\documentclass[prd,twocolumn,nopacs,floatfix,amsmath,nofootinbib,amssymb,floatfix]{revtex4}
\usepackage{graphicx,color,dcolumn,booktabs,bm}
\usepackage{longtable,lscape}
\usepackage{pdfpages}
\usepackage{txfonts}
\usepackage{bbm}
\usepackage{overpic}
\usepackage{amssymb}
\usepackage{indentfirst}
\usepackage{feynmf}   
\usepackage{slashed}  
\usepackage{cases}
\usepackage{color}
\usepackage{multirow}
\usepackage{threeparttable}
\usepackage{epstopdf}
\usepackage{enumerate}
\usepackage{graphicx,color,dcolumn,booktabs,bm}
\usepackage[colorlinks, citecolor=blue,anchorcolor=red,menucolor=red, linkcolor=red,filecolor=red,urlcolor=blue,frenchlinks=red]{hyperref}

\graphicspath{{Figures/}} %

\begin{document}

\title{$b$-hadron spectroscopy study based on the similarity of double bottom baryon and bottom meson}
\author{Bing Chen$^{1,3,4}$}\email{chenbing@ahstu.edu.cn}
\author{Si-Qiang Luo$^{2,3,4}$}\email{luosq15@lzu.edu.cn}
\author{Ke-Wei Wei$^{5}$}\email{weikw@hotmail.com}
\author{Xiang Liu$^{2,3,4}$\footnote{Corresponding author}}\email{xiangliu@lzu.edu.cn}
\affiliation{$^1$School of Electrical and Electronic Engineering, Anhui Science and Technology University, Bengbu 233000, China\\
$^2$School of Physical Science and Technology, Lanzhou University, Lanzhou 730000, China\\
$^3$Lanzhou Center for Theoretical Physics, Key Laboratory of Theoretical Physics of Gansu Province, and Frontiers Science Center for Rare Isotopes, Lanzhou University, Lanzhou 730000, China\\
$^4$Research Center for Hadron and CSR Physics, Lanzhou University $\&$ Institute of Modern Physics of CAS, Lanzhou 730000, China\\
$^5$College of Science, Henan University of Engineering, Zhengzhou 451191, China}

\date{\today}

\begin{abstract}

The dynamical similarity which exists between the $\lambda$-mode excited $bbq$ baryons ($q$ refers to the $u$, $d$, and $s$ quarks) and the $\bar{b}q$ mesons inspires us to carry out a combined study of their spectroscopy. In this work, the masses and strong decays of these low-lying $b\bar{q}$ and $bbq$ states are studied by the same theoretical methods, and the dynamical similarity which is implied in their mass spectra and strong decays are also discussed. The recent discovered $\bar{b}q$ states, including the $B_J(5840)$, $B_J(5970)$, $B_{sJ}(6064)$, and $B_{sJ}(6114)$, are analyzed. According to our result, the $B_J(5840)$ could be assigned as a 2$S$ state, while the $B_J(5970)$ could be regarded as a member of the $1D(2^-,~3^-)_{j_q=5/2}$ doublet. The $B_{sJ}(6064)$ and $B_{sJ}(6114)$ are probably the $D$-wave states. Especially, they could be explained as the members of the $1D(1^-,~2^-)_{j_q=3/2}$ and $1D(2^-,~3^-)_{j_q=5/2}$ doublets, respectively. The predicted masses and decay properties of other unknown $\bar{b}q/bbq$ states may provide useful clues to the future experiment.

\end{abstract}

 \maketitle


\section{Introduction}\label{sec1}

The study of hadron spectroscopy is a useful approach to deepen our understanding of the nonperturbative behavior of strong interaction. In the past decade, big progress has been made on the observation of bottom baryons, which is a crucial step to construct a complete $b$-hadron spectroscopy.
Among these observations, the $\Lambda_b(5912)$,  $\Lambda_b(5920)$, $\Lambda_b(6072)$, $\Lambda_b(6146)$, $\Lambda_b(6152)$, $\Xi_b(6100)$, $\Xi_b(6327)$, and $\Xi_b(6333)$ which were reported in Refs. \cite{LHCb:2012kxf,CDF:2013pvu,CMS:2020zzv,LHCb:2020lzx,CMS:2021rvl,LHCb:2021ssn} have a close relation to
the low-lying $\Lambda_b$ and $\Xi_b$ baryons. Meanwhile, these $P$-wave $\Sigma_b$, $\Xi_b^\prime$, and $\Omega_b$ states were also established one by one which are mainly due to the effort from LHCb~\cite{LHCb:2018haf,LHCb:2018vuc,LHCb:2020tqd,LHCb:2020xpu}. In contrast to the situation of single bottom baryons, none of the double bottom baryons has been discovered, which is becoming not only a challenge but also an opportunity for the future experiments.
If further checking the status of bottom and bottom-strange mesons, we can find that the families of bottom and bottom-strange mesons are far from being established since there are only seven states collected in each of the bottom and bottom-strange meson families, which can be referred to the Particle Data Group (PDG) \cite{ParticleDataGroup:2020ssz}. Thus, constructing the spectroscopy of bottom and bottom-strange mesons is still on the way.


Before introducing the motivation of the present work, we should mention the lesson from the investigation of single heavy baryons.
Lanzhou group has systematically studied the single heavy baryons by the nonrelativistic constituent quark models~\cite{Chen:2016iyi,Chen:2017aqm,Chen:2017gnu,Chen:2018orb,Chen:2018vuc,Chen:2019ywy}, where the diquark picture was employed. These studies not only depicted the newly observed single heavy baryons, but also have stood the test of later experimental observations. For examples, the predicted properties of $P$-wave $\Omega_b$~\cite{Chen:2018vuc} and $D$-wave $\Xi_b$~\cite{Chen:2019ywy} states were confirmed by the LHCb experiments~\cite{LHCb:2020tqd,LHCb:2021ssn}. To some extent, our works could be regarded as a typical example of applying the diquark picture to hadron spectroscopy.

\begin{figure}[htbp]
\begin{center}
\includegraphics[width=6.8cm,keepaspectratio]{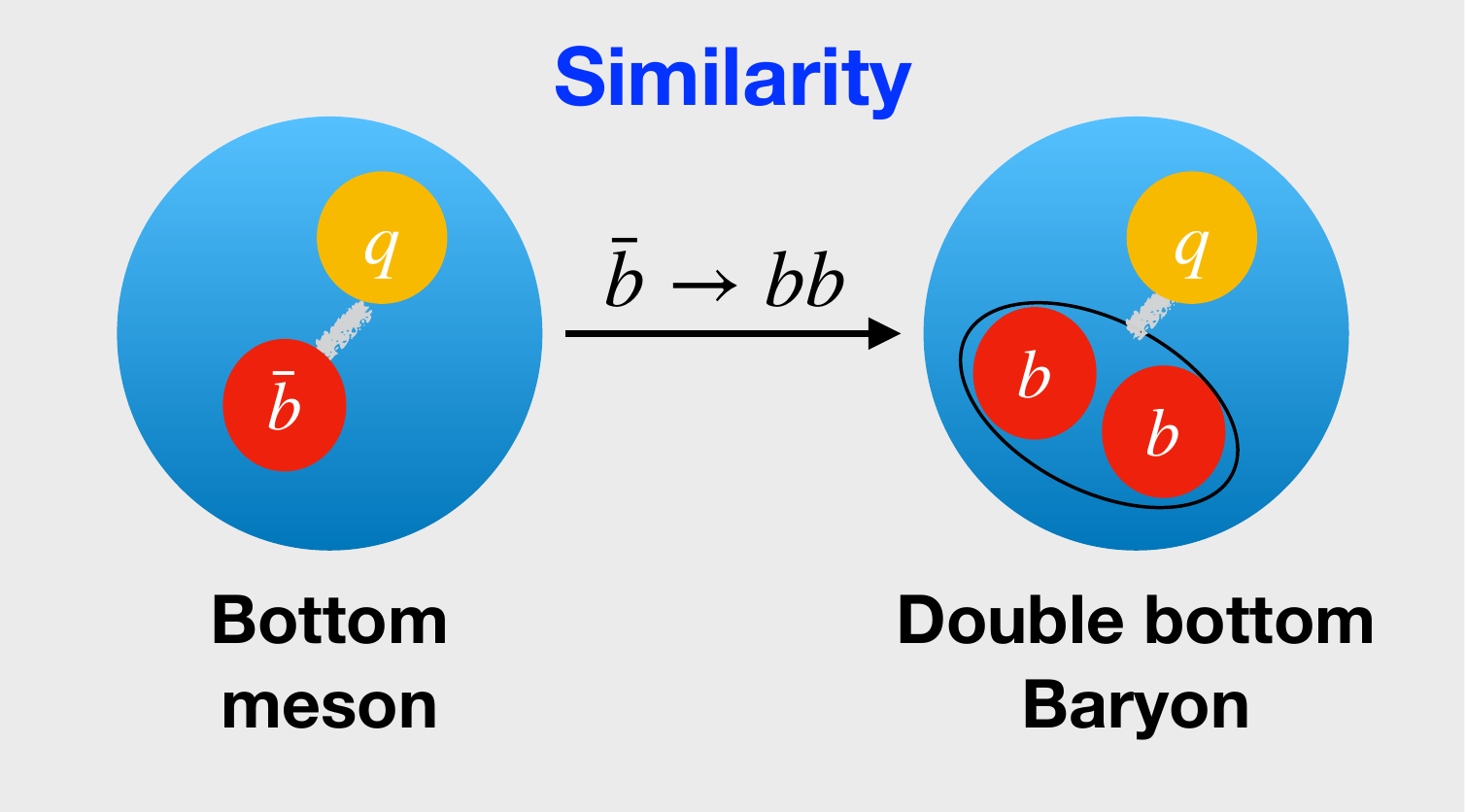}
\caption{The similarity between the bottom meson and the double bottom baryon.}\label{Fig1}
\end{center}
\end{figure}

For the double bottom baryon, we may also treat this interesting system as a quasi-two-body problem if putting two bottom quarks together which is also a typical diquark. Indeed, the heavy-diquark-light-quark picture has been taken into account when calculating the mass spectrum of double bottom baryons \cite{Gershtein:2000nx,Kiselev:2002iy,Ebert:2002ig}.
In the following, we should mention the similarity between the double bottom baryon and the
bottom meson if the diquark picture is considered. As illustrated in Fig. \ref{Fig1}.
the double bottom baryon system could be simplified as a quasi-two-body system in the diquark picture, where the heavy diquark has the same color
structure as a heavy antiquark in the bottom meson. Thus, it is naturally expect that the $\lambda$-mode excited $bbq$ baryons have the similar dynamics to the bottom mesons. So it provides a possibility to carry a combined study of the mass spectrum of these two kinds of $b$-hadrons.

\begin{figure}[htbp]
\begin{center}
\includegraphics[width=7.8cm,keepaspectratio]{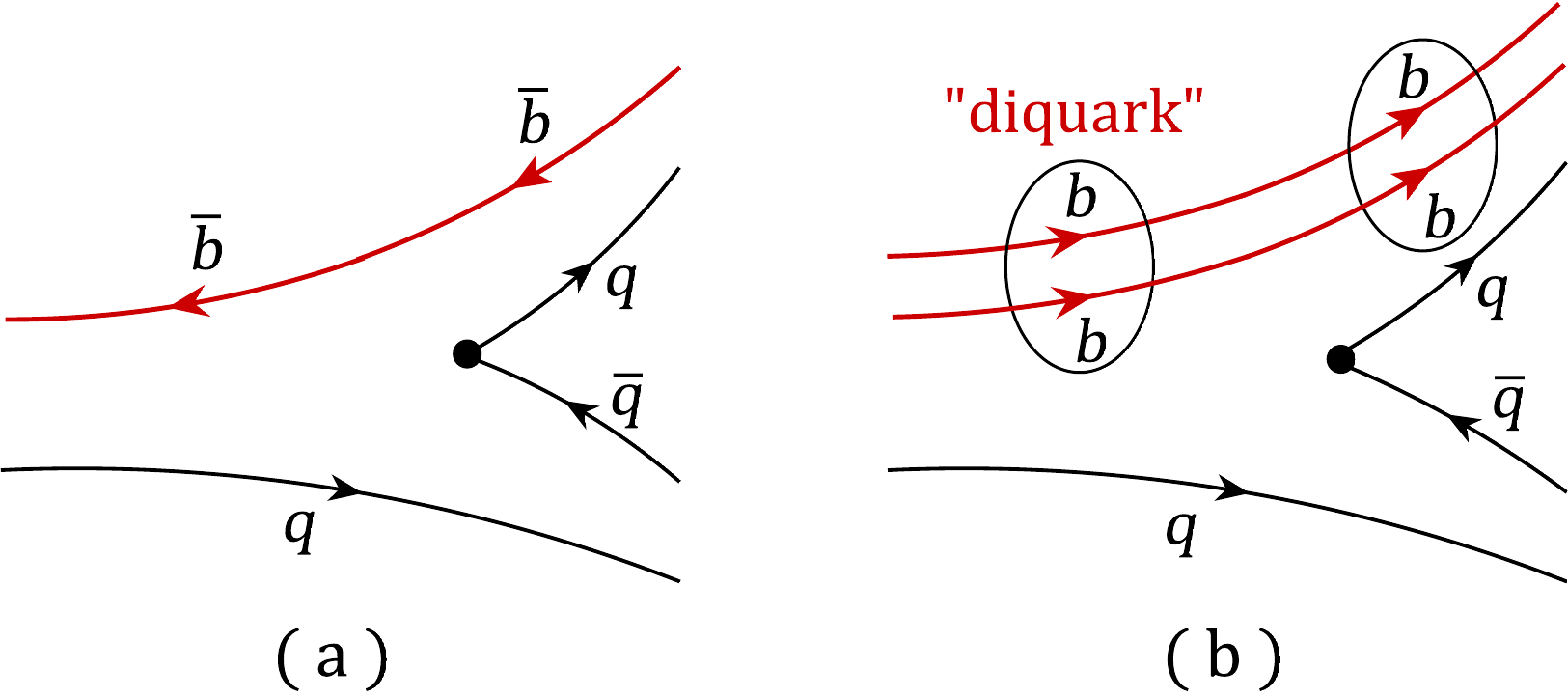}
\caption{(a) A topological diagram for a decay process of an excited $\bar{b}q$ state;  (b) A diagram for a decay process of a $bbq$ excitation.}\label{Fig2}
\end{center}
\end{figure}

Besides giving their mass spectrum, in this work, we also focus on the strong decay properties of the bottom meson and double bottom baryon together.
As shown later, most of the low-excited $\lambda$-mode $bbq$ states are expected to be below the $\Lambda_bB$ and $\Xi_bB$ thresholds. Thus, for the strong decays of these double bottom baryon excitations, two bottom quarks which transit into a final $bbq$ state could be treated as a whole (see Fig. \ref{Fig2}). So the heavy diquark is just a spectator in the decay process, which is similar to the strong decay of an excited $\bar{b}q$ meson.
Therefore, the similarity should exist not only in the spectroscopy of low-lying $\bar{b}q$ and $bbq$ states, but also in their strong decays. It means that these two kinds of $b$-hadrons can be investigated in the same theoretical scheme for their masses and strong decays. In this way, the model parameters fitted by the known $\bar{b}q$ mesons can provide a valuable reference to the $bbq$ system since none of $bbq$ states has been discovered.  The proposed approach is different from most of studies in the literature, where the $\bar{b}q$ and $bbq$ states were studied individually.

The paper is organized as follows. After the introduction, the theoretical scheme is introduced in details in Sec. \ref{sec2}. The mass spectra and strong decay behaviors of these low-lying $\bar{b}q$ and $bbq$ states are investigated in Sec. \ref{sec3} and \ref{sec4}, respectively. The similarities of the $\bar{b}q$ and $bbq$ systems which are implied in the predicted masses and strong decays are further discussed in Sec. \ref{sec5}. Finally, the paper ends with the conclusion and outlook in Sec. \ref{sec6}.


\section{Theoretical scheme}\label{sec2}

It has been a long time since Savage and Wise proposed the superflavor symmetry which can be related to the discussion of the properties of the $\bar{Q}q$ mesons and $QQq$ baryons~\cite{Savage:1990di}. The emergence of superflavor symmetry is a consequence of the heavy quark limit. In the limit of $m_Q\rightarrow\infty$, two heavy quarks in the $QQq$ baryonic system may form a small weakly bound color triplet subsystem. In literature~\cite{White:1991hz}, the $\{QQ\}$ subsystem is also referred to be the heavy diquark. As an approximation, this heavy diquark could be regarded as a static source of color, which plays essentially the same role as the heavy antiquark in a heavy-light meson.

The superflavor symmetry has been developed and applied by Cohen~\cite{Cohen:2006jg}, Roberts~\cite{Eakins:2012jk,Eakins:2012fq}, Ma~\cite{Ma:2017nik} and their collaborators. An important application of the superflavor symmetry is to obtain the properties of these unknown double heavy baryons from the well-measured heavy-light mesons. The superflavor symmetry was used in Ref.~\cite{Cohen:2006jg} to predict the mass difference of two ground double heavy baryons with $J^P=1/2^+$ and $J^P=3/2^+$, respectively, where the masses of two ground heavy-light mesons (the pseudoscalar and vector mesons) were taken as the input. Based on the superflavor symmetry, the selection rule and the spin-counting relation for the strong decays of the double heavy baryon were studied in Ref.~\cite{Eakins:2012fq}. The superflavor symmetry was incorporated into an effective Lagrangian approach in Ref.~\cite{Ma:2017nik}. Then the masses and decays of ground and 1$P$ (with $J^P=1/2^-$ and $J^P=3/2^-$) doubly heavy baryons were predicted. Recently, the superflavor symmetry was also extended to investigate the tetraquark system $Q_iQ_j\bar{q}_k\bar{q}_l$~\cite{Eichten:2017ffp}.


In this work, we are also dedicated to the application of superflavor symmetry. Differently, we shall take the concrete dynamical model to systematically investigate the mass spectrum and decay behavior of these low-lying $\bar{Q}q$ mesons and $QQq$ baryons together, where the superflavor symmetry will be considered. Concretely, a nonrelativistic quark potential model\footnote{In principle, the dynamics of light quark $q$ in the $\bar{Q}q$ and $QQq$ systems should be depicted by the relativistic models. When one tries to calculate the masses of low-lying hadron states, however, it is believed that a nonrelativistic model could include many relativistic corrections by a redefinition of its parameters~\cite{Mezoir:2008vx}, such as the constituent quark masses, the effective parameters in the static potential [see Eq.~(\ref{eq1})]. So it can explain why the mass spectra obtained by the relativized quark model are qualitatively similar to those of the usual nonrelativistic model~\cite{Capstick:1986ter}. For the doubly heavy baryons, the excited energies given by the relativistic quark model~\cite{Ebert:2002ig} and the nonrelativistic quark model~\cite{Gershtein:2000nx,Kiselev:2002iy} are also nearly equal.} is employed to calculate the mass spectra of the $\bar{b}q$ mesonic and $bbq$ baryonic states. Here, we take the Cornell potential~\cite{Eichten:1978tg} to phenomenologically depict the confining interaction between a $\bar{3}$ color component\footnote{It corresponds to the $\{bb\}$ diquark in a $bbq$ baryon state or the antiquark $\bar{b}$ in a $\bar{b}q$ meson state, which is denoted by the symbol $\mathbbm{h}$ in the following discussion.} and the light quark. Based on this consideration, we may construct the Schr\"{o}dinger equation for the discussed system, i.e.,
\begin{equation}
\left(-\frac{\nabla^2}{2m_\mu}-\frac{4\alpha}{3r}+br-C+\frac{32\alpha\sigma^3e^{-\sigma^2r^2}}{9\sqrt{\pi}m_{\mathbbm{h}}m_q}\textbf{s}_{\mathbbm{h}}\cdot\textbf{s}_{q} \right)\psi_{nL} = E\psi_{nL}. \label{eq1}
\end{equation}
Here, $m_{\mathbbm{h}}$ and $\textbf{s}_{\mathbbm{h}}$ denote the mass and the spin of the heavy component $\mathbbm{h}$ in the $\bar{b}q/bbq$ hadrons. Equation~(\ref{eq1}) has included the spin-spin contact hyperfine interaction between the heavy component $\mathbbm{h}$ and the light quark. For the $\Xi_{bb}$ and $\Omega_{bb}$ baryons $s_{\mathbbm{h}}=1$ is determined, while for the $B$ and $B_s$ mesons, we take $s_{\mathbbm{h}}=1/2$. The parameters $\alpha$, $b$, and $C$ stand for the strength of the color Coulomb potential, the strength of linear confinement, and a mass-renormalized constant, respectively. By solving the Schr\"{o}dinger equation, the average masses of the $bbq$ and $\bar{b}q$ hadrons are obtained. When the following spin-dependent interactions are further incorporated, we can further obtain the mass of these discussed $\bar{b}q/bbq$ states. The first spin-dependent interaction
\begin{equation}
H_{\textup{T}}=\frac{4\alpha_s}{3m_{\mathbbm{h}}m_q}\frac{1}{r^3}\left(\frac{3\textbf{s}_{\mathbbm{h}}\cdot\emph{\textbf{r}}~\textbf{s}_q\cdot\emph{\textbf{r}}}{r^2}-\textbf{s}_{\mathbbm{h}}\cdot\textbf{s}_q\right), \label{eq2}
\end{equation}
is a tensor term for depicting the magnetic-dipole-magnetic-dipole color hyperfine interaction. The second spin-dependent term is the spin-orbit interaction
\begin{eqnarray}
H_{\textup{SO}}&=&\left[\left(\frac{2\alpha}{3r^3}-\frac{b}{2r}\right)\frac{1}{m_{\mathbbm{h}}^2}+\frac{4\alpha}{3r^3}\frac{1}{m_{\mathbbm{h}}m_q}\right]\textbf{s}_{\mathbbm{h}}\cdot\emph{\textbf{L}}\nonumber\\&&
+\left[\left(\frac{2\alpha}{3r^3}-\frac{b}{2r}\right)\frac{1}{m_q^2}+\frac{4\alpha}{3r^3}\frac{1}{m_{\mathbbm{h}}m_q}\right]\textbf{s}_q\cdot\emph{\textbf{L}}, \label{eq3}
\end{eqnarray}
which arises from both the short-range one-gluon exchange contribution and the long-range Thomas-precession term.

As shown in Eqs.~(\ref{eq2})-(\ref{eq3}), the ``$\textbf{s}_{\mathbbm{h}}\cdot\emph{\textbf{L}}$'' coupling and the tensor interaction will vanish in the limit of $m_{\mathbbm{h}}\rightarrow\infty$. Thus, the spin-dependent interactions which are denoted in Eqs.~(\ref{eq2})-(\ref{eq3}) could be further expressed as
\begin{equation}
\begin{aligned}\label{eq4}
H_{\textup{SD}}=~&\frac{1}{m_q^2}\left(\frac{2\alpha}{3r^3}-\frac{b}{2r}\right)\textbf{s}_q\cdot\emph{\textbf{L}}+\frac{1}{m_{\mathbbm{h}}m_q}\frac{4\alpha}{3r^3}\left(\emph{\textbf{S}}\cdot\emph{\textbf{L}}+\hat{s}_{\textup{T}}\right)\\
                 &+\frac{1}{m_{\mathbbm{h}}^2}\left(\frac{2\alpha}{3r^3}-\frac{b}{2r}\right)\textbf{s}_{\mathbbm{h}}\cdot\emph{\textbf{L}},
\end{aligned}
\end{equation}
where $\hat{s}_{\textup{T}}$ and $\emph{\textbf{S}}$ are defined as $\hat{s}_{\textup{T}}=(3\textbf{s}_{\mathbbm{h}}\cdot\emph{\textbf{r}}~\textbf{s}_q\cdot\emph{\textbf{r}})/r^2-\textbf{s}_{\mathbbm{h}}\cdot\textbf{s}_q$ and  $\emph{\textbf{S}}=\textbf{s}_q+\textbf{s}_{\mathbbm{h}}$, respectively. For the $\bar{b}q$ and $bbq$ hadrons, the $m_{\mathbbm{h}}$ is much larger than the $m_q$ $(m_{\mathbbm{h}}\gg{m_q})$. Then the first term in the right-hand side of Eq.~(\ref{eq4}) is predominant for the $\bar{b}q/bbq$ systems.

\begin{figure}[htbp]
\begin{center}
\includegraphics[width=8.6cm,keepaspectratio]{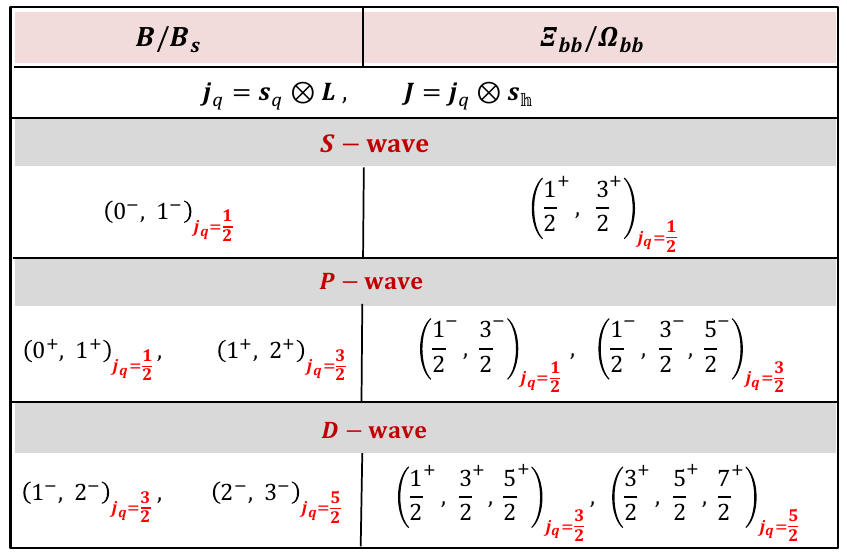}
\caption{The multiplets of low-lying $\bar{b}q/bbq$ states. }\label{Fig3}
\end{center}
\end{figure}

Accordingly, the practical calculation of mass matrix is often performed in the $jj$ coupling scheme, where the basis is defined as $|s_q, L, j_q, s_{\mathbbm{h}}, J\rangle$ with $\emph{\textbf{j}}_q=\textbf{s}_q+\emph{\textbf{L}}$ and $\emph{\textbf{J}}=\emph{\textbf{j}}_q+\textbf{s}_{\mathbbm{h}}$. For the sake of convenience, the basis $|s_q, L, j_q, s_{\mathbbm{h}}, J\rangle$ could be abbreviated as $|j_q, J^P\rangle$ without any confusions. The parities of the $\bar{b}q$ meson and $\lambda$-mode excited $bbq$ baryon are fixed as $P=(-1)^{L+1}$ and $P=(-1)^L$, respectively. As an example, the basis of $D$-wave $bbq$ baryons could be defined as the $|3/2, 1/2^+\rangle$, $|3/2, 3/2^+\rangle$, $|3/2, 5/2^+\rangle$, $|5/2, 3/2^+\rangle$, $|5/2, 5/2^+\rangle$, and $|5/2, 7/2^+\rangle$, which could be grouped into the $j_q=3/2$ and $j_q=5/2$ triplets. The multiplets of all low-lying $\bar{b}q/bbq$ states are presented in Fig. \ref{Fig3}.

In reality, the masses of the antiquark $\bar{b}$ and the diquark $\{bb\}$ are finite. Then, the degeneracy of the states in a multiplet is broken, and there exist mixing of the states with same $J^P$ in the different multiplets. Due to $m_{\mathbbm{h}}\gg{m_q}$, however, the mixing effect of the $\bar{b}q/bbq$ states with the same $J^P$ is not obvious. It means that the basis $|j_q, J^P\rangle$ defined in the heavy quark limit could represent the physical state approximatively.

The superflavor symmetry is reflected not only in the mass spectrum of the $\bar{b}q$ and $bbq$ systems, but also in their strong decay behavior. Since the dynamics of light quark $q$ in the $\bar{b}q$ and $bbq$ systems is independent of the spin and mass of the heavy component $\mathbbm{h}$ and only the light quark $q$ takes an active part in the strong decay process, the states of the corresponding multiplets (see Fig. \ref{Fig3}) with the same $j_q$ but different $J^P$ should have the similar decay behavior. We may take the $P$-wave $\bar{b}q$ and $bbq$ states as an example. These four $P$-wave $\bar{b}q$ mesons can be arranged into two doublets, i.e., the $(0^+, 1^+)_{j_q=1/2}$ and $(1^+, 2^+)_{j_q=3/2}$, while these five $P$-wave $bbq$ baryons could be grouped in a doublet and a triplet, i.e., the $(1/2^-, 3/2^-)_{j_q=1/2}$ doublet and the $(1/2^-, 3/2^-, 5/2^-)_{j_q=3/2}$ triplet. Since the $\bar{b}q$ mesons in the $(1^+, 2^+)_{j_q=3/2}$ doublet have been measured to be the narrow states~\cite{ParticleDataGroup:2020ssz}, the superflavor symmetry indicates that the $P$-wave $\Xi_{bb}$ and $\Omega_{bb}$ baryons in the $(1/2^-, 3/2^-, 5/2^-)_{j_q=3/2}$ triplet should also be the narrow states. This phenomenon can be reflected by the practical calculations which will be presented in Sec. \ref{sec4}.


In this work, it is reasonable to extend the decay formula which was proposed by Eichten, Hill, and Quigg in Ref.~\cite{Eichten:1993ub} to study the decay of the double bottom baryons. This strong decay formula (or named as the EHQ formula) incorporates the heavy quark symmetry for depicting a decay process of an excited heavy-light meson~\cite{Isgur:1991wq}. The EHQ formula, which has been used to explain the decays of the charm mesons and charm baryons in our previous works~\cite{Chen:2015lpa,Chen:2017aqm}, could be improved as
\begin{equation}
\Gamma^{A\rightarrow BC}_{j_C,\ell} = \xi\,\left|\mathcal
{C}^{s_{\mathbbm{h}},j_B,J_B}_{j_C,j_A,J_A}\,\mathcal
{M}^{j_A,j_B}_{j_C,\ell}(p/\beta)\right|^2 \,p\,
e^{-p^2/6\beta^2}. \label{eq5}
\end{equation}
In this way, the EHQ formula can be used to study the decay properties of both the bottom mesons and the $\lambda$-mode excited double bottom baryons. The only difference of the $bbq$ and $\bar{b}q$ states is that the spin of the diquark $\{bb\}$ in a $bbq$ baryon is 1 $(s_{\{bb\}}=1)$, while the spin of the antiquark $\bar{b}$ in a $\bar{b}q$ meson is $1/2$ $(s_{\bar{b}}=1/2)$. Then, the spin of the heavy quark $s_Q$ in the original EHQ formula~\cite{Eichten:1993ub} has been replaced by $s_{\mathbbm{h}}$ in Eq.~(\ref{eq5}).

\emph{A} and $B$ in Eq.~(\ref{eq5}) represent the initial and final heavy-light hadrons, respectively, and $C$ denotes the light flavor meson. The magnitude of three-momentum for a final state is denoted as $p$ in the rest frame of the initial state. The flavor factor $\xi$ in Eq.~(\ref{eq5}) has been given in Ref.~\cite{Chen:2012zk}. The symbols $s_C$ and $\ell$ represent the spin of the light hadron $C$ and the orbital angular momentum relative to $B$, respectively. The normalized coefficient $\mathcal {C}^{s_Q,j_B,J_B}_{j_C,j_A,J_A}$ is rewritten as
\begin{eqnarray}\label{eq6}
\begin{split}
\mathcal {C}^{s_{\mathbbm{h}},j_B,J_B}_{j_C,j_A,J_A}=(-1)^{J_A+j_B+j_C+s_{\mathbbm{h}}}~&\sqrt{(2j_A+1)(2J_B+1)}\\ &\times\left\{
           \begin{array}{ccc}
                    s_{\mathbbm{h}}  & j_B & J_B\\
                    j_C              & J_A & j_A\\
                    \end{array}
     \right\}.
\end{split}
\end{eqnarray}
where $\emph{\textbf{j}}_C \equiv \emph{\textbf{s}}_C + \vec{\ell}$. The $\mathcal {C}^{s_Q,j_B,J_B}_{j_C,j_A,J_A}$ denoted by Eq. (\ref{eq6}) reflects the requirement of the heavy quark symmetry. The transition factors $\mathcal {M}^{j_A,j_B}_{j_C,\ell}(p/\beta)$,  which is relevant to the nonperturbative dynamics, could be obtained by the various phenomenological models. In our previous work~\cite{Chen:2015lpa,Chen:2017aqm}, the transition factors were extracted by the $^3P_0$ model~\cite{Micu:1968mk,LeYaouanc:1972vsx,LeYaouanc:1988fx}. Here, we directly borrow the transition factors from Ref.~\cite{Chen:2015lpa} for the calculation. The parameter $\beta$ denotes the scale of harmonic oscillator wave function for the hadrons involved in the discussed transitions. Since the $bbq$ baryon has been simplified as a quasi-two-body system in the diquark picture, the strong decays of the low-excited $bbq$ and $\bar{b}q$ states have the same $\mathcal {M}^{j_A,j_B}_{j_C,\ell}(p/\beta)$. This is nothing but the consequence of superflavor symmetry which has been stressed above.

The parameters involved in the adopted potential model are collected in Table~\ref{table2}. The dimensionless parameter $\gamma$ in the $^3P_0$ model was fixed as 0.125 by the decay width of the $P$-wave $D_2(2640)^0$ state \cite{Chen:2015lpa}, where the value of $\beta$ is taken as 0.38 GeV. One may notice that the mass of $bb$ diquark in Table \ref{table1} is more than twice of the antiquark $\bar{b}$. According to the results in Ref.~\cite{Karliner:2014gca}, the effective mass of $b$ quark which was fixed by the discovered bottom baryons is larger than the mass of antiquark $\bar{b}$ in the bottom mesons. So it implies that the constituent quark mass of $b$ quark in a bottom baryon could be different from the mass of $\bar{b}$ in a bottom meson. In our calculation, the mass of antiquark $\bar{b}$ is fixed as 4.64 GeV by the measured bottom mesons. On the other hand, the mass of $\{bb\}$ diquark is taken as 9.55 GeV, which is larger than the mass of $\Upsilon$ state ($m_{\Upsilon}=9.46$ GeV). This is consistent with the expectation of quark potential models. In fact, the masses of lowest $\{bb\}$ diquark (the $1^3S_1$ state) in Refs.~\cite{Gershtein:2000nx,Kiselev:2002iy,Ebert:2002ig} were also larger than the mass of $\Upsilon$ state.

With these parameters as the input, the mass spectrum, the decay width, and the corresponding branching ratio of the discussed $\bar{b}q$ and $bbq$ hadrons are obtained in the following sections.

\begin{table}[htbp]
\caption{The values of the parameters for the $\bar{b}q$ and $bbq$ states in the nonrelativistic quark potential model. Here, the parameter $\sigma$  is taken as 1.10 GeV for all kinds of hadrons. $m_{\mathbbm{h}}$ refers to the mass of heavy component in the $\bar{b}q$ and $bbq$ hadrons.
}\label{table1}
\renewcommand\arraystretch{1.3}
\begin{tabular*}{86mm}{c@{\extracolsep{\fill}}ccccc}
\toprule[1pt]\toprule[1pt]
 Parameters   & $m_{q}$ (GeV)     & $m_{\mathbbm{h}}$ (GeV)     & $\alpha$      & $b$  (GeV$^2$)    & $C$ (GeV)  \\
\toprule[1pt]
$B$           & 0.45              & 4.64                        & 0.50          & 0.138             & 0.135  \\
$B_s$         & 0.54              & 4.64                        & 0.50          & 0.138             & 0.077 \\
$\Xi_{bb}$    & 0.45              & 9.55                        & 0.42          & 0.130             & 0.190  \\
$\Omega_{bb}$ & 0.54              & 9.55                        & 0.42          & 0.130             & 0.130  \\
\bottomrule[1pt]\bottomrule[1pt]
\end{tabular*}
\end{table}

\section{Bottom and bottom-strange mesons}\label{sec3}

\subsection{The $B$ meson}


\begin{table}[htbp]
\caption{These observed $B$ and $B_s$ mesons~\cite{LHCb:2020pet,ParticleDataGroup:2020ssz}. Here, the candidates of the 1$S$ and 1$P$ states of the $B/B_s$ meson are listed in the first and second rows, while the possible candidates for the $2S$ or $1D$ states are collected in the last row.}\label{table2}
\renewcommand\arraystretch{1.2}
\begin{tabular*}{85mm}{c@{\extracolsep{\fill}}c }
\toprule[1pt]\toprule[1pt]
      $B$     &     $B_s$       \\
\toprule[1pt]
   $B(5280)/B^\ast(5325)$         & $B_s(5367)/B_s^\ast(5416)$            \\
  $B_J^\ast(5732)/B_1(5721)/B_2^\ast(5747)$     &  $B_{sJ}^\ast(5850)/B_{s1}(5830)/B_{s2}^\ast(5840)$           \\
   $B_J(5840)/B_J(5970)$     & $B_{sJ}(6064)/B_{sJ}(6114)$           \\
\bottomrule[1pt]\bottomrule[1pt]
\end{tabular*}
\end{table}

As shown in Table \ref{table2}, the low-lying $B$ mesons are far from being well established. At present, only the $B(5280)$, $B^\ast(5325)$, $B_1(5721)$, $B_2^\ast(5747)$, $B_J^\ast(5732)$, $B_J(5840)$, and $B_J(5970)$ are collected by the PDG~\cite{ParticleDataGroup:2020ssz}. Among them, the $B(5280)$, $B^\ast(5325)$, $B_1(5721)$, and $B_2^\ast(5747)$ were established in experiment without controversy. However, two broad 1$P$ bottom mesons are not established. As a disputed candidate of the 1$P$ $B$ meson, the $B_J^\ast(5732)$ was reported in Refs.~\cite{OPAL:1994hqv,DELPHI:1994fnu,ALEPH:1998unp,L3:1999pdo}, where the measured resonant parameters from
different experiments are listed in Table \ref{table3}.

Exploring higher excited $B$ mesons was continuing. The L3 Collaboration reported a bottom meson which could be a 2$S$ or 1$D$ candidate in the hadronic decay process of the $Z$ boson~\cite{L3:1999pdo}. Its mass and decay width were measured to be $5937\pm21\pm4$ MeV and $50\pm22\pm5$ MeV, respectively.
However, this $B$ state has never been confirmed by other experiments and none of other the higher excited $B$ state was found in the next many years. In 2013, the CDF Collaboration found a state $B(5970)$ in the $B\pi$ final states~\cite{CDF:2013www}. Two years later, the LHCb Collaboration reported two $B$ resonances, the $B_J(5840)$ and $B_J(5960)$~\cite{LHCb:2015aaf}. So far, the spin-parity quantum numbers of these reported $B$ mesons are still undetermined (see the discussions in Ref.~\cite{li:2021hss}). In this work, we try to identify their properties by combing our theoretical result with the experimental data.

\begin{table}[htbp]
\caption{The resonance parameters (in MeV) of $B_J^\ast(5732)$ state
from different collaborations.
}\label{table3}
\renewcommand\arraystretch{1.3}
\begin{tabular*}{86mm}{c@{\extracolsep{\fill}}ccc}
\toprule[1pt]\toprule[1pt]
 Mass               & Decay width       & Collaboration                & Year       \\
\toprule[1pt]
 5681$\pm$11        & 116$\pm$24        & OPAL~\cite{OPAL:1994hqv}     & 1995    \\
 5732$\pm$5$\pm$20  & 145$\pm$28        & DELPHI~\cite{DELPHI:1994fnu} & 1995     \\
 5695$^{+17}_{-19}$ & 53$^{+26}_{-19}$  & ALEPH~\cite{ALEPH:1998unp}   & 1998     \\
 5670$\pm$10$\pm$23 & 70$\pm$21$\pm$25  & L3~\cite{L3:1999pdo}         & 1999    \\
\bottomrule[1pt]\bottomrule[1pt]
\end{tabular*}
\end{table}

\begin{table*}[htbp]
\caption{A comparison of our predicted masses of the $B$ and $B_s$ mesons with other results Refs.~\cite{DiPierro:2001dwf,Ebert:2009ua,Godfrey:2016nwn,Asghar:2018tha} and the experimental data~\cite{LHCb:2020pet,ParticleDataGroup:2020ssz} (in MeV).} \label{table4}
\renewcommand\arraystretch{1.2}
\begin{tabular*}{176mm}{l@{\extracolsep{\fill}}cccccccccccc}
\toprule[1pt]\toprule[1pt]
   & \multicolumn{6}{c}{Bottom meson}  &   \multicolumn{6}{c}{Bottom-strange meson}    \\
\cline{2-7}\cline{8-13}
State  &    Expt.  &  Our  &    Ref.~\cite{DiPierro:2001dwf} & Ref.~\cite{Ebert:2009ua}  & Ref.~\cite{Godfrey:2016nwn} &  Ref.~\cite{Asghar:2018tha}  & Expt.  &  Our  &    Ref.~\cite{DiPierro:2001dwf} & Ref.~\cite{Ebert:2009ua}  & Ref.~\cite{Godfrey:2016nwn} &  Ref.~\cite{Asghar:2018tha} \\
\midrule[1pt]
 $1^1S_0$        &  5280  & 5279 & 5279  & 5280 & 5312 & 5268 & 5367 & 5368 & 5373 & 5372 & 5394 & 5377 \\
 $1^3S_1$        &  5325  & 5325 & 5324  & 5326 & 5371 & 5329 & 5416 & 5414 & 5421 & 5414 & 5450 & 5422 \\
 $2^1S_0$        &  5889  & 5903 & 5886  & 5890 & 5904 & 5877 &      & 5977 & 5985 & 5976 & 5984 & 5929 \\
 $2^3S_1$        &        & 5929 & 5920  & 5906 & 5933 & 5905 &      & 6001 & 6019 & 5992 & 6012 & 5949 \\
 $3^1S_0$        &        & 6357 & 6320  & 6379 & 6335 & 6288 &      & 6413 & 6421 & 6467 & 6410 & 6305 \\
 $3^3S_1$        &        & 6378 & 6347  & 6387 & 6355 & 6308 &      & 6431 & 6449 & 6475 & 6429 & 6319 \\
 $1^3P_0$        &        & 5724 & 5706  & 5749 & 5756 & 5704 &      & 5795 & 5804 & 5833 & 5831 & 5770 \\
 $1P_1^{\prime}$ &        & 5759 & 5742  & 5774 & 5784 & 5739 &      & 5834 & 5842 & 5865 & 5861 & 5801 \\
 $1P_1$          &  5726  & 5726 & 5700  & 5723 & 5777 & 5755 & 5829 & 5819 & 5805 & 5831 & 5857 & 5803 \\
 $1^3P_2$        &  5740  & 5740 & 5714  & 5741 & 5797 & 5769 & 5840 & 5833 & 5820 & 5842 & 5876 & 5822 \\
 $2^3P_0$        &        & 6181 & 6163  & 6221 & 6213 & 6129 &      & 6236 & 6264 & 6318 & 6279 & 6160 \\
 $2P_1^{\prime}$ &        & 6218 & 6194  & 6281 & 6228 & 6161 &      & 6268 & 6296 & 6345 & 6296 & 6186 \\
 $2P_1$          &        & 6202 & 6175  & 6209 & 6197 & 6175 &      & 6278 & 6278 & 6321 & 6279 & 6196 \\
 $2^3P_2$        &        & 6220 & 6188  & 6260 & 6213 & 6190 &      & 6289 & 6292 & 6359 & 6295 & 6208 \\
 $1^3D_1$        &        & 6099 & 6025  & 6119 & 6110 & 6022 & 6114 & 6153 & 6127 & 6209 & 6182 & 6057 \\
 $1D_2^{\prime}$ &        & 6110 & 6037  & 6121 & 6124 & 6026 &      & 6164 & 6140 & 6218 & 6196 & 6059 \\
 $1D_2$          &        & 5989 & 5985  & 6103 & 6095 & 6031 &      & 6080 & 6095 & 6189 & 6169 & 6064 \\
 $1^3D_3$        &  5994  & 5996 & 5993  & 6091 & 6106 & 6031 & 6064 & 6086 & 6103 & 6191 & 6179 & 6063 \\
 $1^3F_2$        &        & 6376 & 6264  & 6412 & 6387 & 6259 &      & 6415 & 6369 & 6501 & 6454 & 6273 \\
 $1F_3^{\prime}$ &        & 6382 & 6271  & 6420 & 6396 & 6249 &      & 6421 & 6376 & 6515 & 6462 & 6277 \\
 $1F_3$          &        & 6223 & 6220  & 6391 & 6358 & 6264 &      & 6307 & 6332 & 6468 & 6425 & 6265 \\
 $1^3F_4$        &        & 6226 & 6226  & 6380 & 6364 & 6252 &      & 6311 & 6337 & 6475 & 6432 & 6267 \\
\bottomrule[1pt]\bottomrule[1pt]
\end{tabular*}
\end{table*}


\subsubsection{The $1P$ states}

\begin{table}[htbp]
\caption{The strong decay widths of the $1P$ states of the $B$ meson (in MeV). Here, the forbidden decay channel is denoted by ``$\times$''} \label{table5}
\renewcommand\arraystretch{1.2}
\begin{tabular*}{86mm}{@{\extracolsep{\fill}}lcccc}
\toprule[1pt]\toprule[1pt]
Decay     &  $1^3P_0$        &  $1P_1^\prime$  &   $1P_1$     &   $1^3P_2$    \\
       \cline{2-3}\cline{4-5}
modes     & $B(5724)$        & $B(5759)$       &  $B_1(5726)$ &   $B(5740)$    \\
\midrule[0.8pt]
 $B\pi$          &  255.9    & $\times$        & $\times$     & 14.1           \\
 $B^\ast\pi$     & $\times$  & 245.1           & 30.9         & 12.6           \\
 Total           &  255.6    & 245.1           & 30.9         & 26.7          \\
 Expt.~\cite{LHCb:2015aaf}   &                 &              & 30.1$\pm$5.0 & 24.5$\pm$2.5   \\
\bottomrule[1pt]\bottomrule[1pt]
\end{tabular*}
\end{table}


The $1P$ bottom mesons $B_1(5721)$ and $B_2^\ast(5747)$ have been established by experiments~\cite{ParticleDataGroup:2020ssz}.
As shown in Table~\ref{table4}, the masses of the $B_1(5721)$ and $B_2^\ast(5747)$
can be well reproduced in our calculation, where we assign these two states as the $1P_1$ and $1^3P_2$ states of the $B$ meson. Thus, the reported $B_1(5721)$ and $B_2^\ast(5747)$ belong to the $1P(1^+,~2^+)_{j_q=3/2}$ doublet, which indicates that the $B_1(5721)$ and $B_2^\ast(5747)$ should have similar decay behavior.
By the EHQ formula, the total decay width of the $B(1^3P_2)$ meson is calculated to be 26.7 MeV, which is consistent with the experimental data~\cite{LHCb:2015aaf}. Under this assignment, the partial width ratio of the $B\pi$ and $B^\ast\pi$ decay mode is obtained to be
\begin{equation}
R(B_2^\ast(5747)^0)=\frac{\mathcal{B}(B_2^\ast(5747)^0\rightarrow{B^\ast\pi})}{\mathcal{B}(B_2^\ast(5747)^0\rightarrow{B\pi})}=0.89, \label{eq7}
\end{equation}
which is also comparable with the LHCb result for the $B_2^\ast(5747)$ state,  i.e., $R(B_2^\ast(5747)^0)=0.71\pm0.14\pm0.30$~\cite{LHCb:2015aaf}. Thus, the study of the decay behavior of the $B_2^\ast(5747)$ enforces the conclusion of the $B_2^\ast(5747)$ as a $B(1^3P_2)$ meson.
If assigning the $B_1(5726)$ to be a pure $1P_1$ meson with $j_q=3/2$, its total decay width is predicted to be 17.8 MeV in the heavy quark limit, which indicates the $B_1(5726)$ to be a narrow state. In reality, the mass of the $b$ quark is not infinite, which makes that the heavy quark symmetry is broken slightly. Thus, the $B_1(5726)$ state  may contain a small $1P(1/2,~1^+)$ component, i.e., two $1P_1^\prime$ and $1P_1$ states of the $B$ meson listed in Table~\ref{table4} should be as the mixtures between the $1P(1/2,~1^+)$ and $1P(3/2,~1^+)$ states
\begin{eqnarray}\label{eq8}
\begin{aligned}
 \left(
           \begin{array}{c}
                     B^\prime_1(5759)\\
                     B_1(5726)\\
                    \end{array}
     \right)&=\left(
           \begin{array}{rr}
                    \cos\theta_{1P}    & -\sin\theta_{1P} \\
                    \sin\theta_{1P}    & \cos\theta_{1P} \\
                    \end{array}
     \right)  \left(
           \begin{array}{c}
                     |1^+,~1/2\rangle \\
                     |1^+,~3/2\rangle \\
                    \end{array}
     \right).
\end{aligned}
\end{eqnarray}
The mixing angle $\theta$ is fixed to be $-175.3^\circ$ by the nonrelativistic quark potential model. Thus, the physical state $B_1(5726)$ has the dominant component of the $1P(3/2,~1^+)$ state, while the undiscovered $B^\prime_1(5759)$ is predominantly the $1P(1/2,~1^+)$ state. When considering this mixing effect, the decay width of the $B_1(5726)$ increases to 30.9 MeV as shown in Table \ref{table5}, which is consistent with the LHCb result~\cite{LHCb:2015aaf}. In a word, it is reasonable to categorize the $B_1(5721)$ and $B_2^\ast(5747)$ as the $P$-wave bottom mesons.

We also investigated the other two $1P$ states of the $B$ meson. Here, the masses of the $1^3P_0$ and $1P_1^\prime$ states of the $B$ meson are predicted to be 5724 MeV and 5759 MeV, respectively, which are comparable with the former theoretical results reported in Refs.~\cite{DiPierro:2001dwf,Kher:2017mky,Ebert:2009ua,Godfrey:2016nwn,Asghar:2018tha}. As shown in Table~\ref{table5}, the  total decay widths of the $1^3P_0$ and $1P_1^\prime$ states of the $B$ meson are given as 255.6 MeV and 245.1 MeV, respectively, in line with expectations of the $B(1^3P_0)$ and $B(1P_1^\prime)$ states having the broad widths as suggested by other theoretical approaches like the $^3P_0$ model~\cite{Sun:2014wea}, the chiral quark model~\cite{Zhong:2008kd}, and the QCD sum rule~\cite{Zhu:1998wy}. Since the $B(1^3P_0)$ and $B(1P_1^\prime)$ states have the broad widths around 200 MeV, experimentally identifying these two $P$-wave states is not an easy task, which may naturally explain why two members in the $1P(0^+,~1^+)_{1/2}$ doublet were not yet established in experiment. At present, it is too early to assign the $B_J^\ast(5732)$ state, which was observed by the OPAL~\cite{OPAL:1994hqv}, DELPHI~\cite{DELPHI:1994fnu}, ALEPH~\cite{ALEPH:1998unp}, and L3~\cite{L3:1999pdo} collaborations, to a member of the $1P(0^+,~1^+)_{1/2}$ doublet before achieving more precise data.

\subsubsection{The $2S$ states}


Precisely measuring the resonance parameters of $2S$ heavy-light mesons is full of challenges. We may take the $2S$ charmed mesons as an example to illustrate this point. As the 2$S$ states of $D$ meson, the $D_0(2550)^0$~\cite{BaBar:2010zpy,LHCb:2013jjb,LHCb:2019juy} and $D_1^\ast(2600)^0$~~\cite{BaBar:2010zpy,LHCb:2013jjb,LHCb:2019juy,LHCb:2016lxy} have been reported for many years. However, the measurements based on the different production processes gave the different results for the resonance parameters (see the review paper~\cite{HFLAV:2019otj} for more details). The main reason is that distinguishing the broad resonances $D_0(2550)^0$ and $D_1^\ast(2600)^0$ in the $D^\ast\pi$ invariant mass spectrum is difficult\footnote{The broad states $D_0(2550)^0$ and $D_1^\ast(2600)^0$ may decay into the same decay channel $D^\ast\pi$. Another Okubo-Zweig-Iizuka (OZI) allowed  decay channel of the $D_1^\ast(2600)^0$ is the $D\pi$ mode.}. A similar situation can happen for the 2$S$ states of the bottom meson (see the predicted widths listed in Table~\ref{table6}). And, a low-energy photon from the $B^\ast\to{B\gamma}$ decay was not reconstructed in the realistic analysis of the CDF~\cite{CDF:2013www} and LHCb~\cite{LHCb:2015aaf} experiments, which makes the results of the $B_J(5840)$ and $B_J(5970)$ to be more uncertain. Thus, the precise measurements are desired for the 2$S$ $B$ mesons in the future.
In this work, the following ratio of the partial widths of the $B\pi$ and $B^\ast\pi$ channels
\begin{equation}
R[B^\ast(2^3S_1)]=\frac{\mathcal{B}(B^\ast(2^3S_1)\rightarrow{B^\ast\pi})}{\mathcal{B}(B^\ast(2^3S_1)\rightarrow{B\pi})}=1.94, \label{eq9}
\end{equation}
is predicted for the $B(2^3S_1)$ meson, which can be examined in the future experiment.

If comparing our result with the measurement of the $B_J(5840)$ and $B_J(5970)$, we find that the $B_J(5840)$ could be interpreted as a good candidate of the $2S$ state, while the possibility of the $B_J(5970)$ as an 2$S$ state could be preliminarily excluded due to its relatively narrow width. In fact, the mass difference between the $B_J(5840)$ and the $B_J(5970)$ also disfavors the assignment of the $B_J(5970)$ as an $2S$ candidate when the $B_J(5840)$ has been assigned as a $2S$ candidate (see the obtained mass splitting of two $2S$ $B$ mesons in Table \ref{table4}).

\begin{table}[htbp]
\caption{The predicted decay properties of the 2$S$ states of the $B$ meson and the measured width of the $B_J(5840)$~\cite{ParticleDataGroup:2020ssz} (in MeV).} \label{table6}
\renewcommand\arraystretch{1.2}
\begin{tabular*}{86mm}{@{\extracolsep{\fill}}lcclcc}
\toprule[1pt]\toprule[1pt]
Decay     &  $2^1S_0$        &  $2^3S_1$       &    &              &        \\
          \cline{2-3}
modes     & $B(5903)$        & $B(5929)$       &    \multicolumn{3}{c}{$(Continued)$}        \\
\midrule[0.8pt]
 $B\pi$          & $\times$  & 62.0            & $B_s^\ast{K}$    & 3.5          & 0.7           \\
 $B^\ast\pi$     & 174.2     & 120.3           & $B(5724)\pi$     & 3.0          & $\times$      \\
 $B\eta$         & $\times$  & 5.0             & $B(5759)\pi$     & $\times$     & 2.2           \\
 $B^\ast\eta$    & 2.9       & 4.6             & $B(5726)\pi$     & $\times$     & 0.2           \\
 $B_sK$          & $\times$  & 2.3             & $B(5740)\pi$     & 0.0          & 0.1           \\
                                                        \cline{4-6}
                 &           &                 &  Total           & 183.6        & 197.4          \\
                 &           &                 &  Expt.           & \multicolumn{2}{c}{ 224$\pm$80~\cite{ParticleDataGroup:2020ssz}}   \\
\bottomrule[1pt]\bottomrule[1pt]
\end{tabular*}
\end{table}


\subsubsection{The $1D$ states}

As shown in Fig. \ref{Fig3}, four 1$D$ states of the $B$ meson could be grouped into two doublets  $1D(1^-,~2^-)_{3/2}$ and $1D(2^-,~3^-)_{5/2}$. The average mass of two members in the $1D(1^-,~2^-)_{3/2}$ doublet is around 6100 MeV, which is about 100 MeV larger than the states in the $1D(2^-,~3^-)_{5/2}$ doublet (see Table \ref{table4}). As shown in Table \ref{table7}, the decay properties of these states in different doublets are quite different, while the states in the same doublet have similar decay behavior. The $1^3D_1$ and $1D_2^\prime$ states are expected to be broad, while the $1^3D_3$ and $1D_2$ states are much narrower. Here, the $1^3D_1$ and $1D_2^\prime$ states decay into the $B^{(\ast)}\pi$ channel through $p$-wave, while the $B^{(\ast)}\pi$ decays of the $1^3D_3$ and $1D_2$ states occur via $f$-wave. In addition, the processes $B(1^3D_1)\to{B(5726)\pi}$ and $B(1D_2^\prime)\to{B(5740)\pi}$ proceed via $s$-wave, while the corresponding strong decays of the $1^3D_3$ and $1D_2$ states occur via $d$-wave. Obviously, the phase space of decay processes for the $1^3D_1$ and $1D_2^\prime$ $B$ states are much larger than that of the $1D_2$ and $1^3D_3$ states.

\begin{table}[htbp]
\caption{The calculated decay properties of the 1$D$ states of the $B$ meson and the measured width of the $B_J(5970)$~\cite{LHCb:2015aaf} (in MeV). If the mass of an initial state is below the threshold of a decay channel, the decay process cannot proceed, which is denoted by ``$-$.''} \label{table7}
\renewcommand\arraystretch{1.2}
\begin{tabular*}{86mm}{@{\extracolsep{\fill}}lcccc}
\toprule[1pt]\toprule[1pt]
Decay     &  $1^3D_1$        &  $1D_2^\prime$  &   $1D_2$     &   $1^3D_3$    \\
       \cline{2-3}\cline{4-5}
modes     & $B(6099)$        & $B(6110)$       &  $B(5989)$   &   $B(5996)$    \\
\midrule[0.8pt]
 $B\pi$          &  39.2    & $\times$         & $\times$     & 14.3           \\
 $B^\ast\pi$     & 22.1     & 64.7             & 21.2         & 12.9           \\
 $B\eta$         &  9.7     & $\times$         & $\times$     & 0.2           \\
 $B^\ast\eta$    & 4.4      & 13.7             & 0.1          & 0.1           \\
 $B_sK$          &  8.0     & $\times$         & $\times$     & 0.1           \\
 $B_s^\ast{K}$   & 3.3      & 10.5             & 0.0          & 0.0           \\
 $B\rho$         &  4.9     & 4.1              & $-$          & $-$           \\
 $B^\ast\rho$    & $-$      & 0.6              & $-$          & $-$           \\
 $B\omega$       &  1.3     & 0.7              & $-$          & $-$           \\
 $B^\ast\omega$  & $-$      & 0.0              & $-$          & $-$           \\
 $B(5724)\pi$    & $\times$ & 0.3              & 0.2          & $\times$      \\
 $B(5759)\pi$    &  0.4     & 0.3              & 0.0          & 0.1           \\
 $B(5726)\pi$    &  187.3   & 1.2              & 0.4          & 0.1           \\
 $B(5740)\pi$    &  1.5     & 188.3            & 0.1          & 0.5           \\
 $B(5903)\pi$    &  0.1     & $\times$         & $\times$     & 0.0           \\
 $B(5929)\pi$    &  0.0     & 0.1              & $-$          & $-$           \\
        \cline{2-3}\cline{4-5}
 Total           &  282.2   & 284.5            & 22.0         & 28.3          \\
 Expt.           &          &                  &            \multicolumn{2}{c}{ 55.9$\pm$16.0~\cite{LHCb:2015aaf}}  \\
\bottomrule[1pt]\bottomrule[1pt]
\end{tabular*}
\end{table}

By comparing the measured resonance parameters with the calculated results in Table \ref{table7}, we find that the $B_J(5970)$ state could be explained as a members of $1D(2^-,~3^-)_{5/2}$ doublet since the measured mass of the $B_J(5970)$ is comparable with the predicted value. Since the photon emitted from the state $B^\ast(5325)$ in the decay process $B_J(5970)\to{B^\ast(5325)+\pi}\to{B\gamma+\pi}$ cannot be reconstructed in experiment~\cite{CDF:2013www,LHCb:2015aaf}, the enhancement structure of the $B_J(5970)$ state may contain the signals of the $1^3D_3$ and $1D_2$ states. Our result naturally explains why the measured width of $B_J(5970)$ is about two times larger than the theoretical result in Table \ref{table7}. In this work, we further give the following ratios of the partial widths of the $B^\ast\pi$ and $B\pi$ decay modes
\begin{equation}
R[B(1^3D_1)]=\frac{\mathcal{B}(B(1^3D_1)\rightarrow{B^\ast\pi})}{\mathcal{B}(B(1^3D_1)\rightarrow{B\pi})}=0.56, \label{eq10}
\end{equation}
and
\begin{equation}
R[B(1^3D_3)]=\frac{\mathcal{B}(B(1^3D_3)\rightarrow{B^\ast\pi})}{\mathcal{B}(B(1^3D_3)\rightarrow{B\pi})}=0.90 \label{eq11}
\end{equation}
for the $1^3D_1$ and $1^3D_3$ $B$ mesons, respectively.

\subsection{The $B_s$ mesons}

The low-lying $B_s$ mesons are also far from being established. Up to now, only the $B_s(5367)$, $B_s^\ast(5415)$, $B_{s1}(5830)$, $B_{s2}^\ast(5840)$, and $B_{sJ}^\ast(5850)$ are collected by the PDG~\cite{ParticleDataGroup:2020ssz}. Among them, the $B_s(5367)$, $B_s^\ast(5415)$ (two 1$S$ states), $B_{s1}(5830)$, and $B_{s2}^\ast(5840)$ (two narrow 1$P$ states in the $1P(1^+,~2^+)_{3/2}$ doublet) have been well established in experiment.
However, more efforts should be paid for establishing two $P$-wave $B_s$ states in the $1P(0^+,~1^+)_{1/2}$ doublet, where assigning the $B_{sJ}^\ast(5850)$ reported by OPAL \cite{OPAL:1994hqv} as a $1P$ state is still open to dispute. Recently, two higher $B_s$ resonance structures, the $B_{sJ}(6064)$ and $B_{sJ}(6114)$, were found by the LHCb Collaboration~\cite{LHCb:2020pet}.
Decoding the properties of these observed $B_{sJ}^\ast(5850)$, $B_{sJ}(6064)$, and $B_{sJ}(6114)$ is a task of this work.


\subsubsection{The $1P$ states}

As shown in Tables \ref{table4} and \ref{table8}, the resonance parameter of $B_{s1}(5830)$ and $B_{s2}^\ast(5840)$ can be reproduced in our theoretical scheme if they are treated as the members of the $1P(1^+,~2^+)_{j_q=3/2}$ doublet. The partial width ratios of the $B^{\ast+}{K^-}$ and $B^+K^-$ decay channels
\begin{equation}
R[B_s^{\ast0}(1^3P_2)]=\frac{\mathcal{B}(B_s^{\ast0}(1^3P_2)\rightarrow{B^{\ast+}{K^-}})}{\mathcal{B}(B_s^{\ast0}(1^3P_2)\rightarrow{B^+K^-})}=8.6\%, \label{eq12}
\end{equation}
is obtained by the EHQ formula, which is comparable with the experimental value $\Gamma(B^{\ast+}{K^-})/\Gamma(B^+K^-)=(9.3\pm1.8)\%$~\cite{ParticleDataGroup:2020ssz}. As a counterpart of the $B_1(5721)^0$ in the $B_s$ sector, the $B_{s1}(5830)$ also contains a small $1P(1/2,~1^+)$ component due to the slight breaking of heavy quark symmetry. Thus, the mixing effect should be considered. If taking the mixing angle of $B_1(5721)$ meson as an input, the total decay width of the $B_{s1}(5830)$ is evaluated to be 1.33 MeV, which is comparable with the experimental data~\cite{CDF:2013www,CMS:2018wcx}.

\begin{table}[htbp]
\caption{The strong decay behavior of the $1P$ states of the $B_s$ meson (in MeV).} \label{table8}
\renewcommand\arraystretch{1.2}
\begin{tabular*}{86mm}{@{\extracolsep{\fill}}lcccc}
\toprule[1pt]\toprule[1pt]
Decay          &  $1^3P_0$    &  $1P_1^\prime$  &   $1P_1$     &   $1^3P_2$    \\
       \cline{2-3}\cline{4-5}
modes          & $B_s(5795)$  & $B_s(5835)$     &  $B_s(5819)$ &   $B_s(5833)$    \\
\midrule[0.8pt]
 $BK$          &  226.8       & $\times$        & $\times$     & 1.52           \\
 $B^\ast{K}$   & $\times$     & 189.7           & 1.33         & 0.11           \\
 Total         &  226.8       & 189.7           & 1.33         & 1.63          \\
 Expt.~\cite{LHCb:2015aaf} &              &                 & 0.5$\pm$0.4  & 1.49$\pm$0.27   \\
\bottomrule[1pt]\bottomrule[1pt]
\end{tabular*}
\end{table}

The properties of the $1^3P_0$ and $1P_1^\prime$ states of the $B_s$ meson are still in debate since no clear signal of these states has been observed in experiment. As shown in Table \ref{table8}, the $1^3P_0$ and $1P_1^\prime$ states of the $B_s$ meson are expected to be broad. The similar results were also obtained by other quenched potential models~\cite{Zhong:2008kd,Sun:2014wea,li:2021hss}.\footnote{The thresholds of the $s$-wave channels $BK$ and $B^\ast{K}$ are about 10$\sim$20 MeV below the predicted bare masses of the $1^3P_0$ and $1P_1^\prime$ $B_s$ states. So the effect of nearby closed channels should be important for the $1^3P_0$ and $1P_1^\prime$ $B_s$ states. In fact, the $1^3P_0$ and $1P_1^\prime$ $B_s$ states were found to be the bound states below the corresponding thresholds when the nontrivial coupled channel effect was considered~\cite{Lang:2015hza,Cheng:2017oqh}. So the experiment may find the $B_s$ states in the $1P(0^+,~1^+)_{1/2}$ doublet through the $B_s\pi^0$ and $B_s^{\ast}\pi^0$ decay channels, respectively, if they are below the $B^{(\ast)}K$ thresholds. Obviously, it is premature to regard the $B_{sJ}^\ast(5850)$ as a candidate of the $1P(0^+,~1^+)_{1/2}$ doublet.}


\subsubsection{The $2S$ states}

The average mass of the $2S$ $B_s$ mesons are calculated to be around 6.0 GeV as given in Table \ref{table4}. Their total decay widths are evaluated to be more than 100 MeV, which are listed in Table \ref{table9}. These results indicate that the $2S$ bottom-strange mesons are two broad resonances.
If roughly comparing the measured mass of the newly observed $B_{sJ}(6064)$ with the predictions of the $2S$ $B_s$ mesons, it seems possible to explain the $B_{sJ}(6064)$ as a $2S$ candidate. However, there still exists a difficulty for this assignment.
The width of the $B_{sJ}(6064)$ was measured to be~\cite{LHCb:2020pet}
\begin{eqnarray*}
\Gamma(B_{sJ}(6064))=26\pm4\rm{(stat)\pm4(syst)~MeV},
\end{eqnarray*}
which is much smaller than the theoretical expectation (see Table \ref{table9}).
The possibility of $B_{sJ}(6114)$ as a $B(2S)$ state can also be excluded since its mass is too heavy to be regarded as the $2S$ candidate. In the future, our experimental colleague should pay more effort to perform the search for the $B_s(2S)$ mesons.

\begin{table}[htbp]
\caption{The decay widths of the $B_s(2S)$ mesons (in MeV).} \label{table9}
\renewcommand\arraystretch{1.2}
\begin{tabular*}{86mm}{@{\extracolsep{\fill}}cccccc}
\toprule[1pt]\toprule[1pt]
  Decay modes       & $BK$      & $B^\ast{K}$ &  $B_s\eta$   & $B_s^\ast\eta$   &   Total    \\
\midrule[0.8pt]
 $B(5977)$          &  $\times$ & 118.5       & $\times$     & $-$              & 118.5        \\
 $B^\ast(6001)$     &  62.8     & 96.4        & 3.0          & 0.3              & 162.5        \\
\bottomrule[1pt]\bottomrule[1pt]
\end{tabular*}
\end{table}

More valuable information of the $B_s(2^3S_1)$ meson can be provided to the further experimental exploration. Based on the partial widths of $BK$ and $B^\ast{K}$ channels in Table \ref{table9}, the following partial width ratios
\begin{equation}
R[B_s^\ast(2^3S_1)]=\frac{\mathcal{B}(B^{\ast}(2^3S_1)\rightarrow{B^{\ast+}{K^-}})}{\mathcal{B}(B^{\ast}(2^3S_1)\rightarrow{B^+K^-})}=1.54, \label{eq13}
\end{equation}
is obtained.


\subsubsection{The $1D$ states}

The masses of the $1^3D_1$ and $1D_2^\prime$ $B_s$ mesons are predicted to be 6153 MeV and 6164 MeV, respectively, while the masses of the $1D_2$ and $1^3D_3$ $B_s$ mesons are expected to be 6080 MeV and 6086 MeV. Mass spectrum analysis supports the $B_{sJ}(6064)$ state as a member in the  $1D(2^-,~3^-)_{5/2}$ doublet and the $B_{sJ}(6114)$ state as a member in the $1D(1^-,~2^-)_{3/2}$ doublet. Besides, we also study the strong decays of the $D$-wave $B_s$ mesons, as listed in Table \ref{table10}. The predicted decay behaviors
may enforce to the assignments of $B_{sJ}(6064)$ and $B_{sJ}(6114)$ above. Concretely, the narrow $B_{sJ}(6064)$ could be regarded a candidate of a $1D_2$ or $1^3D_3$ $B_s$ state, while the $B_{sJ}(6114)$ state could be a $1^3D_1$ or $1D_2^\prime$ $B_s$ meson.

\begin{table}[htbp]
\caption{The partial and total decay widths of the 1$D$ states of the $B_s$ meson (in MeV).} \label{table10}
\renewcommand\arraystretch{1.2}
\begin{tabular*}{86mm}{@{\extracolsep{\fill}}lcccc}
\toprule[1pt]\toprule[1pt]
Decay             &  $1^3D_1$   &  $1D_2^\prime$  &   $1D_2$     &   $1^3D_3$    \\
       \cline{2-3}\cline{4-5}
modes             & $B_s(6153)$ & $B_s(6164)$     &  $B_s(6080)$ &   $B_s(6086)$    \\
\midrule[0.8pt]
 $BK$             &  64.1       & $\times$        & $\times$     & 9.8           \\
 $B^\ast{K}$      &  33.1       & 99.1            & 11.8         & 7.4           \\
 $B_s\eta$        &  12.2       & $\times$        & $\times$     & 0.3           \\
 $B_s^\ast{\eta}$ &   5.2       & 16.3            & 0.2          & 0.1           \\
       \cline{2-3}\cline{4-5}
 Total            &  114.6      & 115.4           & 12.0         & 17.6          \\
 Expt.~\cite{LHCb:2020pet}            & \multicolumn{2}{c}{72$\pm$43} & \multicolumn{2}{c}{ 26$\pm$8}  \\
\bottomrule[1pt]\bottomrule[1pt]
\end{tabular*}
\end{table}

In the future, the experiment may measure the following partial width ratios
\begin{equation}
R[B_s^\ast(1^3D_1)]=\frac{\mathcal{B}(B^{\ast}(1^3D_1)\rightarrow{B^{\ast+}{K^-}})}{\mathcal{B}(B^{\ast}(1^3D_1)\rightarrow{B^+K^-})}=0.52, \label{eq14}
\end{equation}
and
\begin{equation}
R[B_s^\ast(1^3D_3)]=\frac{\mathcal{B}(B^{\ast}(1^3D_3)\rightarrow{B^{\ast+}{K^-}})}{\mathcal{B}(B^{\ast}(1^3D_3)\rightarrow{B^+K^-})}=0.76, \label{eq15}
\end{equation}
to examine the properties of the $1^3D_1$ and $1^3D_3$ $B_s$ states.

\begin{table*}[htbp]
\caption{A comparison of our predicted masses of the $\Xi_{bb}$ and $\Omega_{bb}$ baryons with other approaches~\cite{Lu:2017meb,Kiselev:2002iy,Eakins:2012jk,Ebert:2002ig,Gershtein:2000nx,Yoshida:2015tia,Roberts:2007ni} (in MeV).} \label{table11}
\renewcommand\arraystretch{1.1}
\begin{tabular*}{176mm}{l@{\extracolsep{\fill}}ccccccccccccc}
\toprule[1pt]\toprule[1pt]
\multirow{2}{*}{$nL(j_q,~J^P)~$}   & \multicolumn{6}{c}{$\Xi_{bb}$ baryon}  & ~~~  & \multicolumn{6}{c}{$\Omega_{bb}$ baryon}    \\
\cline{2-7}\cline{9-14}
  &    Our  &  Ref.~\cite{Lu:2017meb}  &    Ref.~\cite{Eakins:2012jk} & Ref.~\cite{Ebert:2002ig}  & Ref.~\cite{Gershtein:2000nx} &  Ref.~\cite{Yoshida:2015tia}  &  & Our  &  Ref.~\cite{Lu:2017meb}  &    Ref.~\cite{Ebert:2009ua} & Ref.~\cite{Kiselev:2002iy}  & Ref.~\cite{Yoshida:2015tia} &  Ref.~\cite{Roberts:2007ni} \\
\midrule[1pt]
 $1S(\frac{1}{2},~\frac{1}{2}^+)$ & 10171  & 10138  & 10322  & 10202  & 10093  & 10314 &  & 10266  & 10230  & 10359  & 10210  & 10447  & 10454  \\
 $1S(\frac{1}{2},~\frac{3}{2}^+)$ & 10195  & 10169  & 10352  & 10237  & 10133  & 10339 &  & 10291  & 10258  & 10389  & 10257  & 10467  & 10486  \\
 $2S(\frac{1}{2},~\frac{1}{2}^+)$ & 10738  & 10662  & 10940  & 10832  &        &       &  & 10816  & 10751  & 10970  &        &        &        \\
 $2S(\frac{1}{2},~\frac{3}{2}^+)$ & 10753  & 10675  & 10972  & 10860  &        &       &  & 10830  & 10763  & 10992  &        &        &        \\
 $1P(\frac{1}{2},~\frac{1}{2}^-)$ & 10593  & 10525  & 10694  & 10632  & 10541  & 10703 &  & 10669  & 10605  & 10771  & 10541  & 10796  &        \\
 $1P(\frac{1}{2},~\frac{3}{2}^-)$ & 10606  & 10526  & 10691  & 10647  & 10578  & 10704 &  & 10681  & 10610  & 10785  & 10567  & 10797  &        \\
 $1P(\frac{3}{2},~\frac{1}{2}^-)$ & 10547  & 10504  & 10691  & 10675  & 10567  & 10740 &  & 10641  & 10591  & 10804  & 10578  & 10803  & 10763  \\
 $1P(\frac{3}{2},~\frac{3}{2}^-)$ & 10561  & 10528  & 10692  & 10694  & 10581  & 10742 &  & 10656  & 10611  & 10821  & 10581  & 10805  & 10765  \\
 $1P(\frac{3}{2},~\frac{5}{2}^-)$ & 10560  & 10547  & 10695  & 10661  & 10580  & 10759 &  & 10655  & 10625  & 10798  & 10580  & 10808  & 10766  \\
 $1D(\frac{3}{2},~\frac{1}{2}^+)$ & 10913  &        & 11011  &        &        &       &  & 10971  &        &        &        &        &        \\
 $1D(\frac{3}{2},~\frac{3}{2}^+)$ & 10918  &        & 11011  &        &        &       &  & 10975  &        &        &        &        &        \\
 $1D(\frac{3}{2},~\frac{5}{2}^+)$ & 10921  &        & 11002  &        &        &       &  & 10979  &        &        &        &        &        \\
 $1D(\frac{5}{2},~\frac{3}{2}^+)$ & 10798  &        & 11011  &        &        &       &  & 10891  &        &        &        &        &        \\
 $1D(\frac{5}{2},~\frac{5}{2}^+)$ & 10803  &        & 11002  &        &        &       &  & 10896  &        &        &        &        &        \\
 $1D(\frac{5}{2},~\frac{7}{2}^+)$ & 10805  &        & 11011  &        &        &       &  & 10898  &        &        &        &        & 11042  \\
\bottomrule[1pt]\bottomrule[1pt]
\end{tabular*}
\end{table*}

\section{double bottom baryons}\label{sec4}

\subsection{The $\Xi_{bb}$ baryons}


\subsubsection{The $nS$ states of the $\Xi_{bb}(nS)$ baryon $(n=1,2)$}

\begin{table}[htbp]
\caption{The decay widths of the $2S$ states of the $\Xi_{bb}$ baryon (in MeV).} \label{table12}
\renewcommand\arraystretch{1.3}
\begin{tabular*}{86mm}{@{\extracolsep{\fill}}lcclcc}
\toprule[1pt]\toprule[1pt]
Decay     &  $2S\left(\frac{1}{2},~\frac{1}{2}^+\right)$  &  $2S\left(\frac{1}{2},~\frac{3}{2}^+\right)$    &    &    &        \\
          \cline{2-3}
modes     & $\Xi_{bb}(10738)$        & $\Xi_{bb}(10753)$       &     \multicolumn{3}{c}{$(Continued)$}       \\
\midrule[0.8pt]
 $\Xi^\prime_{bb}\pi$   & 19.4      & 79.3            & $\Xi_{bb}(10593)\pi$    & 1.2          & $\times$     \\
 $\Xi_{bb}^\ast\pi$     & 148.3     & 95.5            & $\Xi_{bb}(10606)\pi$    & $\times$     & 1.3           \\
 $\Xi^\prime_{bb}\eta$  & 0.2       & 1.6             & $\Xi_{bb}(10547)\pi$    & $\times$     & 0.1           \\
 $\Xi_{bb}^\ast\eta$    & $-$       & 0.3             & $\Xi_{bb}(10561)\pi$    & 0.0          & 0.1           \\
                        &           &                 & $\Xi_{bb}(10560)\pi$    & 0.1          & 0.1           \\
                                                        \cline{4-6}
                 &           &                 &  Total           & 169.2       & 178.3          \\
\bottomrule[1pt]\bottomrule[1pt]
\end{tabular*}
\end{table}

The masses of the ground $\Xi_{bb}$ baryons, including the $\Xi^\prime_{bb}$ ($J^P=1/2^+$) and $\Xi^\ast_{bb}$ ($J^P=3/2^+$) states, have been investigated by the different methods or models (see Refs.~\cite{Wei:2016jyk,Oudichhya:2021yln} and references therein).

With the quark potential model introduced in Sec. \ref{sec2}, the mass of the $\Xi^\prime_{bb}$ state is predicted to be 10171 MeV, which is comparable with the results from Refs.~\cite{Silvestre-Brac:1996myf,Ebert:2002ig,Albertus:2006ya,Karliner:2014gca,Brown:2014ena}.\footnote{For the $QQq$ baryon, there only exists the experimental observation of the double charm baryon $\Xi^\prime_{cc}(3621)$ from the LHCb Collaboration~\cite{LHCb:2017iph,LHCb:2018pcs,LHCb:2019epo}. Before the discovery of $\Xi^\prime_{cc}(3621)$, some theoretical groups successfully
predicted its mass~\cite{Silvestre-Brac:1996myf,Ebert:2002ig,Albertus:2006ya,Karliner:2014gca,Brown:2014ena}. These groups also predicted the $\Xi^\prime_{bb}$ state in the mass range of 10.14$\sim$10.20 GeV. The nearly equal predictions were also achieved for the mass of $\Xi^\prime_{bb}$ state by other approaches including the QCD sum rule~\cite{Wang:2010hs,Wang:2018lhz}, the Salpeter model with AdS/QCD inspired potential~\cite{Giannuzzi:2009gh}, and the extended chromomagnetic model~\cite{Weng:2018mmf}.}
The complete mass spectra of these low-lying $\Xi_{bb}$ states are listed in Table \ref{table11}. The mass difference of the $\Xi^\prime_{bb}(1S)$ and $\Xi^\ast_{bb}(1S)$ states is predicted to be 24 MeV, which is comparable with these results from Refs.~\cite{Ebert:2002ig,Karliner:2014gca,Yoshida:2015tia,Weng:2018mmf}.
There only exists the weak decays for the $\Xi^\prime_{bb}$ state, while the $\Xi^\ast_{bb}$ state can transit into the $\Xi^\prime_{bb}$ state by emitting a photon.

The masses of the $2S$ states of the $\Xi_{bb}$ baryon are predicted to be about 550 MeV above the ground $\Xi_{bb}$ states, which allows the $2S$ $\Xi_{bb}$ states to decay into a ground or 1$P$ $\Xi_{bb}$ state plus a light meson. The strong decay properties of these two $2S$ states of the $\Xi_{bb}$ baryon are given in Table~\ref{table12}. Our result indicates that the $2S$ states of the $\Xi_{bb}$ baryon are two broad resonances, and the $\Xi^\prime_{bb}\pi$ and $\Xi_{bb}^\ast\pi$ are their main decay modes. Finally, we predict the partial width ratios of $\Xi^\prime_{bb}\pi$ and $\Xi_{bb}^\ast\pi$
\begin{equation}
R[\Xi_{bb}(10738)]=\frac{\Gamma(\Xi_{bb}(10738)~\to~\Xi_{bb}^\ast\pi)}{\Gamma(\Xi_{bb}(10738)~\to~\Xi^\prime_{bb}\pi)}=7.64, \label{eq16}
\end{equation}
and
\begin{equation}
R[\Xi_{bb}(10753)]=\frac{\Gamma(\Xi_{bb}(10753)~\to~\Xi_{bb}^\ast\pi)}{\Gamma(\Xi_{bb}(10753)~\to~\Xi^\prime_{bb}\pi)}=1.20, \label{eq17}
\end{equation}
for the $2S$ states  of the $\Xi_{bb}$ baryon, which could be tested in the future experiment.


\subsubsection{The $1P$ state of the $\Xi_{bb}$ baryon}

Different from the case of the $\bar{b}q$ meson, there are five excited states for the $P$-wave $\Xi_{bb}$ baryons. According to their light degrees
of freedom $j_q$, one can categorize these five $1P$ $\Xi_{bb}$ states into one doublet and one triplet, which are denoted as $1P(1/2^-,~3/2^-)_{1/2}$ and $1P(1/2^-,~3/2^-,~3/2^-)_{3/2}$ (see Fig.~\ref{Fig3}). The predicted average mass of 1$P$ $\Xi_{bb}$ baryons is about 380 MeV higher than that of the $1S$ $\Xi_{bb}$states, which agrees with the expectations in Refs.~\cite{Lu:2017meb,Yoshida:2015tia,Soto:2020pfa}. However, the predictions of ``$\bar{M}(1P)-\bar{M}(1S)$'' for the $\Xi_{bb}$ baryons, which were predicted in Refs.~\cite{Ebert:2002ig,Gershtein:2000nx}, are about 50$\sim$70 MeV larger than our result. Then these different expectations should be tested in a future experiment.

The $\Xi^\prime_{bb}\pi$ and $\Xi_{bb}^\ast\pi$ are the main decay channels of the 1$P$ $\Xi_{bb}$ baryons. According to our results in Table \ref{table13}, two states in the $1P(1/2^-,~3/2^-)_{1/2}$ doublet have broad widths, while the states in the $1P(1/2^-,~3/2^-,~3/2^-)_{3/2}$ triplet are much narrow. Thus, finding the $\Xi_{bb}(10547)$, $\Xi_{bb}(10561)$, and $\Xi_{bb}(10560)$ in the decay channels $\Xi^\prime_{bb}\pi$ and $\Xi_{bb}^\ast\pi$
is suggested. Our conclusion for the decay behaviors of the 1$P$ $\Xi_{bb}$ baryons is consistent to the former work~\cite{He:2021iwx}. We further give the following partial width ratios of the $\Xi^\prime_{bb}\pi$ and $\Xi_{bb}^\ast\pi$ channels
\begin{equation}
R[\Xi_{bb}(10561)]=\frac{\Gamma(\Xi_{bb}(10561)~\to~\Xi_{bb}^\ast\pi)}{\Gamma(\Xi_{bb}(10561)~\to~\Xi^\prime_{bb}\pi)}=2.85, \label{eq15}
\end{equation}
and
\begin{equation}
R[\Xi_{bb}(10560)]=\frac{\Gamma(\Xi_{bb}(10560)~\to~\Xi_{bb}^\ast\pi)}{\Gamma(\Xi_{bb}(10560)~\to~\Xi^\prime_{bb}\pi)}=0.62, \label{eq16}
\end{equation}
for the $1P(3/2,~3/2^-)$ and $1P(3/2,~5/2^-)$ states of $\Xi_{bb}$ baryon.

\begin{table}[htbp]
\caption{The decay widths of the $1P$ states of the $\Xi_{bb}$ baryon (in MeV).} \label{table13}
\renewcommand\arraystretch{1.2}
\begin{tabular*}{86mm}{@{\extracolsep{\fill}}lccccc}
\toprule[1pt]\toprule[1pt]
                 & $\Xi_{bb}(10593)$ & $\Xi_{bb}(10606)$ & $\Xi_{bb}(10547)$ & $\Xi_{bb}(10561)$  & $\Xi_{bb}(10560)$  \\
\midrule[0.8pt]
 $\Xi^\prime_{bb}\pi$  &  262.5       & $\times$        & $\times$     & 3.3    & 8.7           \\
 $\Xi^\ast_{bb}\pi$    &  $\times$    & 266.6           & 9.5          & 9.4    & 5.4           \\
 Total                 &  262.5       & 266.6           & 9.5          & 12.7   & 14.1      \\
\bottomrule[1pt]\bottomrule[1pt]
\end{tabular*}
\end{table}

\subsubsection{The $1D$ states of the $\Xi_{bb}$ baryon}

As shown in Fig.~\ref{Fig3}, six $D$-wave $\Xi_{bb}$ states can be grouped into two triplets, which are distinguished by their light degrees of freedom $j_q$. Interestingly, the average mass of these $\Xi_{bb}$ baryons in the $1D(1/2^+,~3/2^+,~5/2^+)_{3/2}$ triplet is about 100 MeV higher than the members in the $1D(3/2^+,~5/2^+,~7/2^+)_{5/2}$ triplet. Furthermore, the mass differences of these states in the same triplet are only several MeV. Accordingly, the mixing effect of the $1D$ $\Xi_{bb}$ baryons which have the same $J^P$ is not obvious in the $jj$ coupling scheme. Here, the mixing angle of two $J^P=3/2^+$ states fixed by the quark potential model is no more than one degree. There exist the similar result for the case of two $J^P=5/2^+$ $\Xi_{bb}$ states.

\begin{table*}[htbp]
\caption{The decay widths of the $1D$ states of the $\Xi_{bb}$ baryon (in MeV). The $P$-wave $\Xi_{bb}$ baryon, as a daughter state in the decay process, is denoted as $\Xi^{(j_q)}_{bb}(J^P)$. The mass of $P$-wave $\Xi_{bb}$ baryon has been given in Table \ref{table11}.} \label{table14}
\renewcommand\arraystretch{1.3}
\begin{tabular*}{176mm}{@{\extracolsep{\fill}}lcccccccccccccc}
\toprule[1pt]\toprule[1pt]
State          & $\Xi^\prime_{bb}\pi$  & $\Xi^\ast_{bb}\pi$  &  $\Xi^\prime_{bb}\eta$ &  $\Xi^\ast_{bb}\eta$  &  $\Omega^\prime_{bb}K$  &   $\Omega^\ast_{bb}K$  &   $\Xi^{(1/2)}_{bb}(\frac{1}{2}^-)\pi$  &  $\Xi^{(1/2)}_{bb}(\frac{3}{2}^-)\pi$ &  $\Xi^{(3/2)}_{bb}(\frac{1}{2}^-)\pi$  &  $\Xi^{(3/2)}_{bb}(\frac{3}{2}^-)\pi$  &  $\Xi^{(3/2)}_{bb}(\frac{5}{2}^-)\pi$  &   $\Xi^\prime_{bb}(2S)\pi$   &   $\Xi^\ast_{bb}(2S)\pi$ &   Total \\
\midrule[0.8pt]
 $\Xi_{bb}(10913)$   & 60.3     & 7.9  & 11.0     & 1.2  & 7.5      & 0.8  & $\times$  & 0.2  & 185.9 &  2.2  &  0.6  & 0.0      & 0.0 & 277.6 \\
 $\Xi_{bb}(10918)$   & 37.3     & 31.2 & 7.0      & 5.1  & 4.8      & 3.2  & 0.1       & 0.2  & 1.5   & 187.4 &  1.7  & 0.0      & 0.0 & 279.7 \\
 $\Xi_{bb}(10921)$   & $\times$ & 69.8 & $\times$ & 11.6 & $\times$ & 7.4  & 0.2       & 0.1  & 0.3   &  1.2  & 188.5 & $\times$ & 0.0 & 279.1 \\
 $\Xi_{bb}(10798)$   & $\times$ & 13.5 & $\times$ & 0.0  & $\times$ & 0.0  & 0.0       & 0.0  & 0.2   &  0.2  & 0.0   & $\times$ & $-$ & 13.9 \\
 $\Xi_{bb}(10803)$   & 4.3      & 10.9 & 0.0      & 0.0  & 0.0      & 0.0  & 0.0       & 0.0  & 0.2   &  0.1  & 0.2   & $-$      & $-$ & 15.7 \\
 $\Xi_{bb}(10805)$   & 10.6     & 6.2  & 0.0      & 0.0  & 0.0      & 0.0  & $\times$  & 0.0  & 0.0   &  0.1  & 0.3   & $-$      & $-$ & 17.2 \\
\bottomrule[1pt]\bottomrule[1pt]
\end{tabular*}
\end{table*}

The small mixing angle of the $1D$ states of the $\Xi_{bb}$ baryon with the same $J^P$ also results in that three members in one triplet have the similar decay widths. As the members of the $1D(1/2^+,~3/2^+,~5/2^+)_{3/2}$ triplet, the $\Xi_{bb}(10913)$,  $\Xi_{bb}(10918)$, and $\Xi_{bb}(10921)$ states are predicted to be very broad.\footnote{According to our result, the $\Xi_{bb}(10913)$ and $\Xi_{bb}(10918)$ states can also decay into the $B\Lambda_b$ channel via the $p$-wave. Since these two $\Xi_{bb}$ states are just above the threshold of the $B\Lambda_b$ channel, the phase spaces of these decay processes are tiny. Thus, the partial decay widths of the $B\Lambda_b$ decay channel are not large for the broad $\Xi_{bb}(10913)$ and $\Xi_{bb}(10918)$ states.}  The $s$-wave decay channels, including the $\Xi^{(3/2)}_{bb}(\frac{1}{2}^-)\pi$, $\Xi^{(3/2)}_{bb}(\frac{3}{2}^-)\pi$, and $\Xi^{(3/2)}_{bb}(\frac{5}{2}^-)\pi$, mainly contribute to the total widths of the $\Xi_{bb}(10913)$, $\Xi_{bb}(10918)$, and $\Xi_{bb}(10921)$ (see Table \ref{table13}).
The $\Xi_{bb}(10798)$, $\Xi_{bb}(10803)$, and $\Xi_{bb}(10805)$ in the $1D(3/2^+,~5/2^+,~7/2^+)_{5/2}$ triplet are much narrower since their total decay widths are around 10$\sim$20 MeV. 


Our study shows that the $\Xi_{bb}(10913)$ could be found by analyzing the $\Xi^\prime_{bb}\pi$ and $\Xi_{bb}(10547)[j_q=\frac{3}{2},J^P=\frac{1}{2}^-]+\pi$ channels. The $\Xi_{bb}(10918)$ could be searched in the $\Xi^\prime_{bb}\pi$, $\Xi^\ast_{bb}\pi$, and $\Xi_{bb}(10561)[j_q=\frac{3}{2},J^P=\frac{3}{2}^-]+\pi$ channels. The $\Xi_{bb}(10921)$ could be observed in its $\Xi^\ast_{bb}\pi$ and $\Xi_{bb}(10560)[j_q=\frac{3}{2},J^P=\frac{5}{2}^-]+\pi$ channels. The $\Xi^\prime_{bb}\pi$ and $\Xi^\ast_{bb}\pi$ are the main decay modes of these narrow states in the $1D(3/2^+,~5/2^+,~7/2^+)_{5/2}$ triplet. Specifically, the $\Xi_{bb}(10798)$ could be detected in the $\Xi^\ast_{bb}\pi$ channel, while the $\Xi_{bb}(10803)$ and $\Xi_{bb}(10805)$ could be found in their $\Xi^\prime_{bb}\pi$ and $\Xi^\ast_{bb}\pi$ decay channels. We also obtain the following ratios
\begin{equation}
\begin{aligned}\label{eq19}
R[\Xi_{bb}(10913)]=~&\frac{\Gamma(\Xi_{bb}(10913)~\to~\Xi_{bb}^\ast\pi)}{\Gamma(\Xi_{bb}(10913)~\to~\Xi^\prime_{bb}\pi)}=0.13,\\
R[\Xi_{bb}(10918)]=~&\frac{\Gamma(\Xi_{bb}(10918)~\to~\Xi_{bb}^\ast\pi)}{\Gamma(\Xi_{bb}(10918)~\to~\Xi^\prime_{bb}\pi)}=0.84,\\
R[\Xi_{bb}(10803)]=~&\frac{\Gamma(\Xi_{bb}(10803)~\to~\Xi_{bb}^\ast\pi)}{\Gamma(\Xi_{bb}(10803)~\to~\Xi^\prime_{bb}\pi)}=2.53,\\
R[\Xi_{bb}(10805)]=~&\frac{\Gamma(\Xi_{bb}(10805)~\to~\Xi_{bb}^\ast\pi)}{\Gamma(\Xi_{bb}(10805)~\to~\Xi^\prime_{bb}\pi)}=0.58,
\end{aligned}
\end{equation}
for the 1$D$ $\Xi_{bb}$ states.


\subsection{The $\Omega_{bb}$ baryons}

\subsubsection{The $nS$ states of the $\Omega_{bb}$ baryon ($n=1,2$)}

\begin{table}[htbp]
\caption{The decay widths of the $2S$ and $1D$ states of the $\Omega_{bb}$ meson (in MeV).} \label{table15}
\renewcommand\arraystretch{1.2}
\begin{tabular*}{86mm}{@{\extracolsep{\fill}}lccccc}
\toprule[1pt]\toprule[1pt]
State        &  $\Xi^\prime_{bb}K$ &  $\Xi^\ast_{bb}K$ &  $\Omega^\prime_{bb}\eta$ &  $\Omega^\ast_{bb}\eta$ &   Total   \\
\midrule[0.8pt]
$\Omega^\prime_{bb}(10816)$    & 16.8     &  109.9    & 0.0      &  $-$          &  126.7           \\
$\Omega^\ast_{bb}(10830)$      & 73.7     &  77.6     & 0.7      &  $-$          &  152.0           \\
 $\Omega_{bb}(10971)$          & 88.3     & 10.9      & 12.2     & 1.3           & 112.7     \\
 $\Omega_{bb}(10975)$          & 55.2     & 43.7      & 7.8      & 5.3           & 112.0      \\
 $\Omega_{bb}(10979)$          & $\times$ & 98.6      & $\times$ & 12.2          & 110.8   \\
 $\Omega_{bb}(10891)$          & $\times$ & 5.4       & $\times$ & 0.0           & 5.4   \\
 $\Omega_{bb}(10896)$          & 2.1      & 4.5       & 0.0      & 0.0           & 6.6       \\
 $\Omega_{bb}(10898)$          & 5.3      & 2.6       & 0.0      & 0.0           & 7.9     \\

\bottomrule[1pt]\bottomrule[1pt]
\end{tabular*}
\end{table}

The experimental measurements~\cite{ParticleDataGroup:2020ssz} indicate that the $B_s$ (or $D_s$) state is about 100 MeV heavier than the corresponding $B$ (or $D$) state. As shown in Table \ref{table11}, the masses of ground $\Omega_{bb}$ states are expected to be about 100 MeV larger than the ground $\Xi_{bb}$ states, which is similar to the case of the heavy-light mesons.
Searching for the $\Omega^\prime_{bb}$ state via its weak decay processes is suggested, while the $\Omega^\ast_{bb}$ can be found in the decay channel $\Omega^\ast_{bb}\to\Omega^\prime_{bb}+\gamma$.

The predicted masses of the $2S$ $\Omega_{bb}$ states are given in Table \ref{table11}, which  are about 540 MeV higher than the average mass of the ground states of the $\Omega_{bb}$ baryon.
Thus, the decay modes $\Xi^\prime_{bb}K$, $\Xi^\ast_{bb}K$, and  $\Omega^\prime_{bb}\eta$ are allowed for the $2S$ states of the $\Omega_{bb}$ baryon. Here, the total decay widths of two $2S$ states of the $\Omega_{bb}$ baryon are predicted to be larger than 100 MeV (see Table \ref{table15}). It is obvious that the discussed $\Omega^\prime_{bb}(10816)$ and $\Omega^\ast_{bb}(10830)$ are two broad resonances, which can mainly decay into the $\Xi^\prime_{bb}K$ and $\Xi^\ast_{bb}K$ channels. Furthermore, the following partial width ratios
\begin{equation}
\begin{aligned}\label{eq20}
R[\Omega_{bb}(10816)]=\frac{\mathcal{B}(\Omega_{bb}(10816)\rightarrow{\Xi^\ast_{bb}K})}{\mathcal{B}(\Omega_{bb}(10816)\rightarrow{\Xi^\prime_{bb}K})}=6.54,\\
R[\Omega_{bb}(10830)]=\frac{\mathcal{B}(\Omega_{bb}(10830)\rightarrow{\Xi^\ast_{bb}K})}{\mathcal{B}(\Omega_{bb}(10830)\rightarrow{\Xi^\prime_{bb}K})}=1.05,
\end{aligned}
\end{equation}
are calculated for the $2S$ states of the $\Omega_{bb}$ baryon.


\subsubsection{The $1P$ states of the $\Omega_{bb}$ baryon}

\begin{table}[htbp]
\caption{The decay widths of the $1P$ states of the $\Omega_{bb}$ baryon (in MeV).} \label{table16}
\renewcommand\arraystretch{1.2}
\begin{tabular*}{86mm}{@{\extracolsep{\fill}}lccccc}
\toprule[1pt]\toprule[1pt]
                 & $\Omega_{bb}(10669)$ & $\Omega_{bb}(10681)$ & $\Omega_{bb}(10641)$ & $\Omega_{bb}(10656)$  & $\Omega_{bb}(10655)$  \\
\midrule[0.8pt]
 $\Xi^\prime_{bb}K$  &  96.2       & $\times$         & $\times$  & $-$    & $-$           \\
 $\Xi^\ast_{bb}K$    &  $\times$   & $-$              & $-$       & $-$    & $-$           \\
 Total               &  96.2       & narrow           & narrow    & narrow & narrow        \\
\bottomrule[1pt]\bottomrule[1pt]
\end{tabular*}
\end{table}

The obtained masses of these five $1P$ $\Omega_{bb}$ states are listed in Table \ref{table11}, which are about 470 MeV above the ground $\Xi_{bb}$ states. Then, the OZI-allowed strong decays are forbidden for most of $1P$ states of the $\Omega_{bb}$ baryon since they are located below the thresholds of the $\Xi_{bb}(1S)K$ channel. As indicated by our results in Table \ref{table16}, only the $1P(1/2^-,~1/2)$ state (the $\Omega_{bb}(10669)$ in Table \ref{table11}) can proceed via its OZI-allowed decay process.\footnote{It should be careful to the conclusion obtained by the quenched quark model, since the bare masses of the $\Omega_{bb}(10669)$ and $\Omega_{bb}(10681)$ states are near to the thresholds of their $s$-wave decay channels $\Xi^\prime_{bb}K$ and $\Xi^\ast_{bb}K$, respectively. So the nontrivial coupled channel effect should be important for the $1P(1/2,~1/2^-)$ and $1P(1/2,~3/2^-)$ $\Omega_{bb}$ states. Further theoretical study on this issue should be paid. As a matter of fact, the masses of the $\Omega_{bb}(10669)$ and $\Omega_{bb}(10681)$ could be shifted down tens MeV due to the coupled channel effect. Thus, the physical $1P(1/2,~1/2^-)$ and $1P(1/2,~3/2^-)$ $\Omega_{bb}$ states would be below the thresholds of their OZI-allowed decay channels, which implies that both $1P(1/2,~1/2^-)$ and $1P(1/2,~3/2^-)$ $\Omega_{bb}$ baryons are the narrow states. In a word, the five 1$P$ $\Omega_{bb}$ baryons are probably the narrow states. This interesting scenario could be tested by the future experiment.}


\subsubsection{The $1D$ states of the $\Omega_{bb}$ baryon}

As shown in Table \ref{table11}, the mass gap between the $1D(1/2^+,~3/2^+,~5/2^+)_{3/2}$ and $1D(3/2^+,~5/2^+,~7/2^+)_{5/2}$ triplets of the $1D$ $\Omega_{bb}$ baryons is predicted to be about 80 MeV.
Similar to the case for the $\Xi_{bb}$ baryon, the mixing angles of the $D$-wave $\Omega_{bb}$ states with the same $J^P$ are very small due to the large mass of heavy diquark $\{bb\}$. In Table \ref{table15}, we collect the predicted decay behaviors of these $1D$ states of the $\Omega_{bb}$ baryon which are also similar to that of the corresponding $\Xi_{bb}$ counterparts. The $\Omega_{bb}(10971)$, $\Omega_{bb}(10975)$, and $\Omega_{bb}(10979)$ in the $1D(1/2^+,~3/2^+,~5/2^+)_{3/2}$ triplet are expected to be the broad resonances, while the $\Omega_{bb}(10891)$, $\Omega_{bb}(10896)$, and $\Omega_{bb}(10898)$ states in another triplet have the narrow total decay widths. The $\Xi^\prime_{bb}K$ and $\Xi^\ast_{bb}K$ are their main decay channels, which could be as the ideal channels to search for these discussed $D$-wave excited $\Omega_{bb}$ states. The following ratios
\begin{equation}
\begin{aligned}\label{eq21}
R[\Omega_{bb}(10971)]=~&\frac{\Gamma(\Omega_{bb}(10971)~\to~\Xi_{bb}^\ast{K})}{\Gamma(\Omega_{bb}(10971)~\to~\Xi^\prime_{bb}{K})}=0.12,\\
R[\Omega_{bb}(10975)]=~&\frac{\Gamma(\Omega_{bb}(10975)~\to~\Xi_{bb}^\ast{K})}{\Gamma(\Omega_{bb}(10975)~\to~\Xi^\prime_{bb}{K})}=0.79,\\
R[\Omega_{bb}(10896)]=~&\frac{\Gamma(\Omega_{bb}(10896)~\to~\Xi_{bb}^\ast{K})}{\Gamma(\Omega_{bb}(10896)~\to~\Xi^\prime_{bb}{K})}=2.14,\\
R[\Omega_{bb}(10898)]=~&\frac{\Gamma(\Omega_{bb}(10898)~\to~\Xi_{bb}^\ast{K})}{\Gamma(\Omega_{bb}(10898)~\to~\Xi^\prime_{bb}{K})}=0.49,
\end{aligned}
\end{equation}
are predicted, which could be applied to further distinguish the different $\Omega_{bb}$ states in the same triplet. The partial width ratios of the $\Omega_{bb}(10971)$, $\Omega_{bb}(10975)$, $\Omega_{bb}(10896)$, and $\Omega_{bb}(10898)$ are comparable with those of $D$-wave $\Xi_{bb}$ states in Eq.~(\ref{eq19}), which may provide a direct evidence to reflect the dynamical similarity of the $\Xi_{bb}$ and $\Omega_{bb}$ systems.

\begin{figure*}[htbp]
\begin{center}
\includegraphics[width=17.5cm,keepaspectratio]{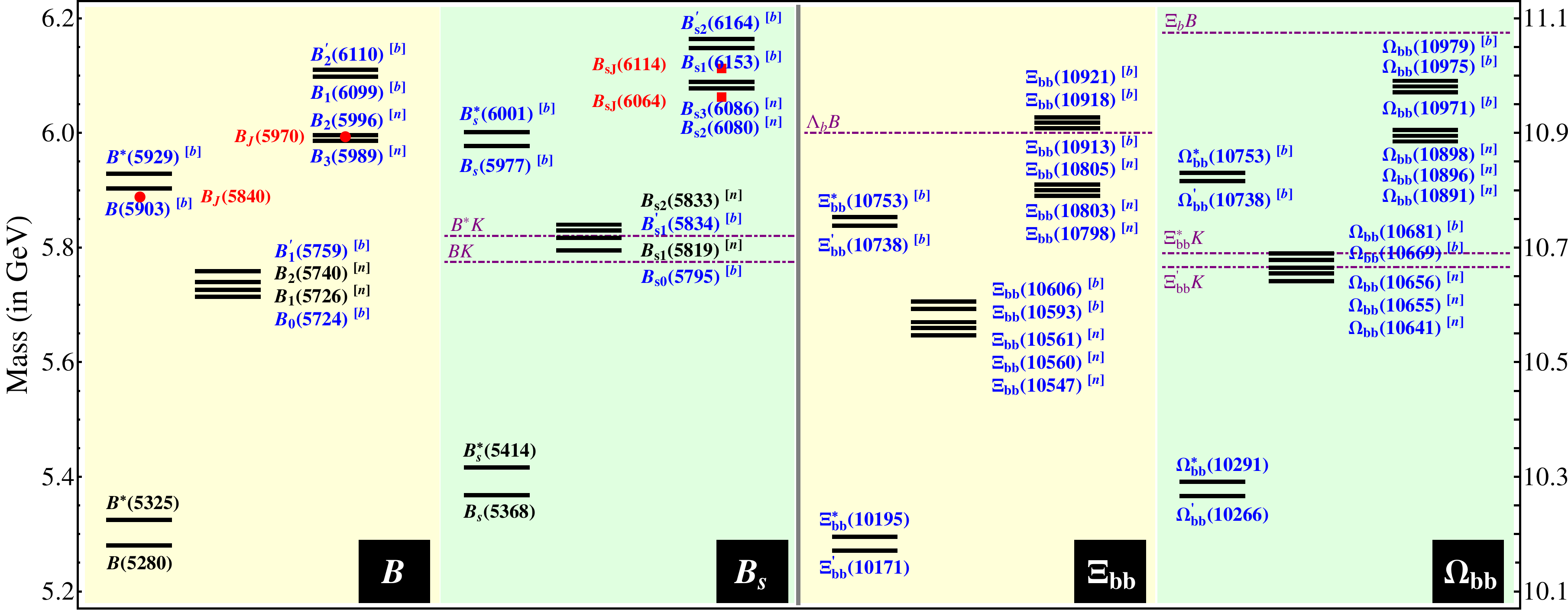}
\caption{The predicted masses and decay properties of these discussed low-lying $B$, $B_s$, $\Xi_{bb}$, and $\Omega_{bb}$ states. The superscripts ``$n$'' and ``$b$'' in the brackets represent the predicted narrow and broad states. The important thresholds of the $B^{(\ast)}K$, $\Lambda_b{B}$, and $\Xi_{bb}^{(\ast)}K$ channels are presented for the $B_s$, $\Xi_{bb}$, and $\Omega_{bb}$ states. The newly observed $B_J(5840)$, $B_J(5970)$, $B_{sJ}(6064)$, and $B_{sJ}(6114)$ are also listed for a comparison.}\label{Fig4}
\end{center}
\end{figure*}


\section{Further discussions of the dynamical similarities between the $\bar{b}q$ and $bbq$ systems}\label{sec5}

For a clarity, the main results for the mass spectrum and strong decay behavior of these discussed low-lying $\bar{b}q$ and $bbq$ states are presented in Fig.~\ref{Fig4}. In the following, we discuss the dynamical similarities between the $\bar{b}q$ and $bbq$ systems behind our results. For the mass spectra, we first define the following ratios
\begin{equation}
\mathcal{R}_1 = \frac{\bar{M}_{2S}-M_{1S}}{\bar{M}_{1P}-\bar{M}_{1S}},~~~~~~~\mathcal{R}_2 = \frac{\bar{M}_{1D}-\bar{M}_{1S}}{\bar{M}_{1P}-\bar{M}_{1S}}. \label{eq22}
\end{equation}
Here, the $\bar{M}_{1S}$, $\bar{M}_{1P}$, $\bar{M}_{2S}$, and $\bar{M}_{1D}$ represent the average masses of the $1S$, $1P$, $2S$, and $1D$ states in the $B$, $B_s$, $\Xi_{bb}$, and $\Omega_{bb}$ sectors. As shown in Table \ref{table17}, we find that $\mathcal{R}_1$ or $\mathcal{R}_2$ for
the $B$, $B_s$, $\Xi_{bb}$, and $\Omega_{bb}$ systems almost have the same values.
Since the 1$P$, 2$S$, and 1$D$ states of the $B/B_s$ mesons are not well established and none of the $bbq$ baryons have been found in experiment, we should take the predicted $\mathcal{R}_1$ and  $\mathcal{R}_2$ of the $\bar{b}q/bbq$ systems to make a comparison with these $\Lambda_b^0$ baryons which could also be regarded as typical heavy-light hadrons in the diquark picture. With the measured masses of $\Lambda_b(5620)$, $\Lambda_b(5912)$, $\Lambda_b(5920)$, $\Lambda_b(6070)$, $\Lambda_b(6146)$, $\Lambda_b(6152)$ states, the $\mathcal{R}_1$ and  $\mathcal{R}_2$ values~\cite{Chen:2021eyk} are also listed in Table \ref{table17}. Obviously, the predicted $\mathcal{R}_1$ or $\mathcal{R}_2$ of $\bar{b}q/bbq$ states are comparable with the measurements of $\Lambda_b$ baryons. The nearly equal values of $\mathcal{R}_1$ or $\mathcal{R}_2$ indicate that the bottom mesons, the single bottom baryons, and the double bottom baryons have the similar dynamics. This novel phenomenon may reflect that the superflavor symmetry is an effective symmetry for these different kinds of bottom hadron systems.


\begin{table}[htbp]
\caption{The predicted ratios of $\mathcal{R}_1$ and  $\mathcal{R}_2$ defined in  Eq.~(\ref{eq22}) for the $\bar{b}q$ and $bbq$ states and a comparison with the measured $\mathcal{R}_1$ and  $\mathcal{R}_2$ of the $\Lambda_b$ baryons.}\label{table17}
\renewcommand\arraystretch{1.2}
\begin{tabular*}{85mm}{c@{\extracolsep{\fill}}ccccc}
\toprule[1pt]\toprule[1pt]
 Ratios           & $\Lambda_b$   & $B$           & $B_s$        & $\Xi_{bb}$    & $\Omega_{bb}$      \\
\toprule[1pt]
$\mathcal{R}_1$   & 1.520         & 1.428         & 1.399        & 1.453         & 1.430           \\
$\mathcal{R}_2$   & 1.780         & 1.699         & 1.680        & 1.715         & 1.702            \\
\bottomrule[1pt]\bottomrule[1pt]
\end{tabular*}
\end{table}

In the following, we should mention the spin-orbit inversion phenomenon of the highly orbital excitations of the $\bar{b}q$ and $bbq$ systems, which may also reflect the dynamical similarity between the $\bar{b}q$ and $bbq$ hadrons.
The spin-orbit inversion of these $P$-wave heavy-light mesons has been discussed in Refs.~\cite{Schnitzer:1978kf,Isgur:1998kr,WooLee:2006kdh} for many years. Since the $P$-wave bottom mesons are not established, the spin-orbit inversion of the $P$-wave heavy-light mesons remains inconclusive.
According to our predictions in Table \ref{table4} and \ref{table11}, the spin-orbit inversion should appear in the $D$-wave $\bar{b}q$ and $bbq$ states since the members in the $j_q=5/2$ multiplet are about 70$\sim$100 MeV lower than the $j_q=3/2$ multiplet (see Fig.~\ref{Fig4}). We try to give a qualitative explanation to this
phenomenon. For the $D$-wave excited $\bar{b}q/bbq$ states, the contribution from the spin-dependent interactions arising from the short-range one-gluon exchange contribution becomes smaller, while the contribution from the Thomas-precession term from the long-range confining potential becomes dominant. Thus, the first term in Eq.~(\ref{eq4}), as the mainly spin-dependent interaction for the orbitally excited $\bar{b}q/bbq$ states, could be a negative number. In fact, the results from Refs.~\cite{DiPierro:2001dwf,Ebert:2009ua,Godfrey:2016nwn} also support the existence of the spin-orbit inversion phenomenon for the $D$-wave $B$ and $B_s$ mesons.

The third dynamical similarity of the focused $\bar{b}q$ and $bbq$ hadron systems can be found in their strong decay behaviors. For the $L$-wave ($L=1,2$) excited $\bar{b}q$ and $bbq$ states, the members in the $j_q=L+\frac{1}{2}$ multiplet seem to be much narrower than those in the $j_q=L-\frac{1}{2}$ multiplet. This phenomenon can be reflected from our results in Tables \ref{table5}, \ref{table7}, \ref{table10}, \ref{table13},  \ref{table14}, and \ref{table15}, where the predicted decay widths of these states in the $j_q=L+\frac{1}{2}$ multiplet are at least one order smaller than those of the states in the $j_q=L-\frac{1}{2}$ multiplet. In fact, this phenomenon has been confirmed by the measured decay widths of the 1$P$ states of the $D$ meson~\cite{ParticleDataGroup:2020ssz}. Here, the decay widths of the $D_1(2420)$ and $D_2^\ast(2460)$, which belong to the $j_q^P=\frac{3}{2}^+$ doublet, were measured to be 25$\sim$50 MeV. In contrast, the decay widths of the $D_0^\ast(2300)$ and $D_1^\prime(2430)$, the states in the $j_q^P=\frac{1}{2}^+$ doublet, are larger than 200 MeV. For enforcing the conclusion for the decay widths of the $j_q=L\pm\frac{1}{2}$ multiplets,
we expect more data of these discussed low-lying $\bar{b}q$ and $bbq$ states to be accumulated in the future experiment.


\section{Conclusion and outlook}\label{sec6}

The dynamical similarities between the $\bar{b}q$ and $bbq$ systems provide us a possibility to carry a combined study of their properties in the same theoretical framework. In this work, we systematically investigate the mass spectra and strong decays of these low-lying $B$, $B_s$, $\Xi_{bb}$, and $\Omega_{bb}$ states. Our result not only decodes these newly observed states $B_J(5840)$, $B_J(5970)$, $B_{sJ}(6064)$, and $B_{sJ}(6114)$, but also reveals the similarities of the $\bar{b}q$ and $bbq$ systems existing in their mass spectra and strong decays.

According to our results, the $B_J(5840)$ could be a 2$S$ state with $J^P=0^-$ or $J^P=1^-$, while the $B_J(5970)$ could be a candidate of the $1D(2^-,~3^-)_{j_q=5/2}$ doublet. Since a low-energy photon from the $B^\ast\to{B\gamma}$ decay was not reconstructed in the realistic measurements of the CDF~\cite{CDF:2013www} and LHCb~\cite{LHCb:2015aaf} collaborations, the $B_J(5840)$ signal can be resulted by two $2S$ states of the $B$ meson. Similar phenomenon may also happen for the $B_J(5970)$ signal which could contain two states of the $1D(2^-,~3^-)_{j_q=5/2}$ doublet. Thus, we suggest the experiment to further identify the spin-parity quantum numbers for these states, or measure the ratio $\Gamma(B^\ast\pi)/\Gamma(B\pi)$, which is helpful to clarify the nature of $B_J(5840)$ and $B_J(5970)$.
By comparing to our predicted masses and decays of the $B_s$ mesons, the $B_{sJ}(6064)$ and $B_{sJ}(6114)$ states which were newly observed by the LHCb Collaboration~\cite{LHCb:2020pet} could be grouped into the $1D(1^-,~2^-)_{j_q=3/2}$ and $1D(2^-,~3^-)_{j_q=5/2}$ doublets, respectively. Experimental measurement of the ratio $\Gamma(B^\ast{K})/\Gamma(B{K})$ for the $B_{sJ}(6064)$ and $B_{sJ}(6114)$ states can further provide the valuable information. For other unknown 1$P$, 2$S$, and 1$D$ $B/B_s$ mesons, we present a complete prediction for their masses and decay properties,  which is useful for the further exploration in the experiment.

The undiscovered $bbq$ baryons are also investigated in this work, where the mass spectra and strong decays of these low-lying $\Xi_{bb}$ and $\Omega_{bb}$ baryons, including the $1S$, $2S$, $1P$, and $1D$ states, are presented. These predictions may provide some clues for the experiment to search for these double bottom baryons in future.

We further point out the similarities between the $\bar{b}q$ and $bbq$ systems which are implied in their mass spectrum and strong decay. As the first similarity of the $\bar{b}q$ and $bbq$ systems, we define two ratios in Eq.~(\ref{eq22}) and show that the predicted $\mathcal{R}_1$ and $\mathcal{R}_2$ values are nearly equal for the $B$, $B_s$ $\Xi_{bb}$, and $\Xi_{bb}$ states (see Table \ref{table17}). We further point out that the predicted $\mathcal{R}_1$ and $\mathcal{R}_2$ of the $\bar{b}q$ and $bbq$ systems are also comparable with the measured result of the $D$ meson. As the second similarity, we find a spin-orbit inversion may occur in the $D$-wave $\bar{b}q$ and $bbq$ states. Finally, we point out that the $\bar{b}q$ and $bbq$ states in the $j_q=L-\frac{1}{2}$ multiplet are much broader than those in the $j_q=L+\frac{1}{2}$ multiplet, which could be regarded as the third similarity of $\bar{b}q$ and $bbq$ systems.

With the accumulation of experimental data, more and more new hadrons have been discovered in the past decades~\cite{Chen:2016qju,Chen:2016spr}. Especially, with the running of LHCb, the studies of heavy flavor hadron enter a new era~\cite{Chen:2021ftn}. When facing on this issue, we have  reasons to believe that more progresses on the heavy flavor hadron will be made in the near future. Surely, the present study may provide some valuable information for the experiments in the next stage.

\section*{Acknowledgements}

This project is supported by the National Natural Science Foundation of China under Grant No. 11305003, No. U1204115, and No. 12047501. X.L. is also supported by the China National Funds for Distinguished Young Scientists under Grant No. 11825503, the National Key Research and Development Program of China under Contract No. 2020YFA0406400.


\end{document}